\documentclass[12pt]{iopart}
\usepackage{amssymb,amsmath}
\usepackage{color} 
\usepackage{accents} 
\usepackage{texdraw}
\usepackage{eucal}
\usepackage{epic,epsfig}
\usepackage{graphicx}
\usepackage[all]{xy}
\usepackage[bf]{caption}

\DeclareMathAccent{\wtilde}{\mathord}{largesymbols}{"65}
\DeclareMathAccent{\what}{\mathord}{largesymbols}{"62}

\def\m@th{\mathsurround=0pt}
\mathchardef\bracell="0365 
\def\upbrall{$\m@th\bracell$}
\def\undertilde#1{\mathop{\vtop{\ialign{##\crcr
    $\hfil\displaystyle{#1}\hfil$\crcr
     \noalign
     {\kern1.5pt\nointerlineskip}
     \upbrall\crcr\noalign{\kern1pt
   }}}}\limits}
\def\theequation{\arabic{section}.\arabic{equation}}
\newcommand{\wb}[1]{\overline{#1}}

\newcommand{\wh}{\widehat}
\newcommand{\wt}{\widetilde}
\newcommand{\ut}{\undertilde}
\def\hypotilde#1#2{\vrule depth #1 pt width 0pt{\smash{{\mathop{#2}
\limits_{\displaystyle\widetilde{}}}}}}
\def\hypohat#1#2{\vrule depth #1 pt width 0pt{\smash{{\mathop{#2}
\limits_{\displaystyle\widehat{}}}}}}

\newcommand{\cQ}{\mathcal{Q}} 
\newcommand{\cH}{\mathcal{H}}
\newcommand{\cA}{\mathcal{A}}

\newcommand{\done}{\\$\square$\par\smallskip\smallskip\noindent\ignorespacesafterend}
\newcommand{\DONE}{\\$\blacksquare$\par\smallskip\smallskip\noindent\ignorespacesafterend}

\newcommand{\ssp}{\mathfrak{p}}

\newcommand{\ssk}{\mathfrak{k}}

\newcommand{\ssl}{\mathfrak{l}}

\newcommand{\ssq}{\mathfrak{q}}

\newcommand{\sst}{\mathfrak{t}}

\newcommand{\bun}{\boldsymbol{1}}

\newcommand{\tbc}{\,^{t\!}\boldsymbol{c}}
\newcommand{\tbu}{\,^{t\!}{\bu}}

\newcommand{\bblu}{\begin{color}{blue}}
\newcommand{\bred}{\begin{color}{red}}
\newcommand{\ecl}{\end{color}}

\newcommand{\bA}{\boldsymbol{A}}

\newcommand{\bK}{\boldsymbol{K}} 
\newcommand{\bL}{\boldsymbol{L}} 
\newcommand{\bM}{\boldsymbol{M}}

\newcommand{\gm}{\gamma}

\newcommand{\dd}{\delta}

\newcommand{\ld}{\lambda}

\newcommand{\be}{\begin{equation}}
\newcommand{\ee}{\end{equation}}
\newcommand{\bea}{\begin{eqnarray}}
\newcommand{\eea}{\end{eqnarray}}
\newcommand{\bse}{\begin{subequations}}
\newcommand{\ese}{\end{subequations}}
\newcommand{\nn}{\nonumber}

\newcommand{\ol}{\overline}

\newcommand{\uo}{\accentset{o}{u}}
\newcommand{\po}{\accentset{o}{p}}
\newcommand{\qo}{\accentset{o}{q}}
\newcommand{\Po}{\accentset{o}{P}}
\newcommand{\Qo}{\accentset{o}{Q}}
\newcommand{\UU}{Z}
\newcommand{\UM}{S}
\newcommand{\UD}[2]{S^{(#1,#2)}}

\newcommand{\bu}{\boldsymbol{u}}

\newcommand{\bbb}{\boldsymbol{b}} 
\newcommand{\bc}{\boldsymbol{c}}

\newcommand{\brr}{\boldsymbol{r}} 

\newcommand{\mbx}{{\boldsymbol x}}
\newcommand{\mby}{{\boldsymbol y}}

\newcommand{\bphi}{{\boldsymbol \phi}}

\begin{document}
\title{Soliton Solutions for ABS Lattice Equations: \\ 
I Cauchy Matrix Approach}  
\author{Frank Nijhoff$^1$, James Atkinson$^2$, Jarmo Hietarinta$^3$} 
\address{
$^1$ Department of Applied Mathematics, University of Leeds, Leeds LS2 9JT, UK \\ 
$^2$ Department of Mathematics and Statistics, Latrobe University, Melbourne, Australia\\ 
$^3$ Department of Physics and Astronomy, University of Turku, FIN-20014 Turku, FINLAND
}
\date{today}
\begin{abstract}
In recent years there have been new insights into the integrability of quadrilateral lattice equations, i.e. partial difference 
equations which are the natural discrete analogues of integrable partial differential equations in 1+1 dimensions. 
In the scalar (i.e. single-field) case there now exist classification results by Adler, Bobenko and 
Suris (ABS) leading to some new examples in addition to the lattice equations ``of KdV type'' that were 
known since the late 1970s and early 1980s. In this paper we review the construction of soliton solutions for the KdV type lattice  
equations and use those results to construct $N$-soliton solutions for all lattice equations in the ABS list except for
the elliptic case of Q4, which is left to a separate treatment. 
 
\end{abstract}

\section{Introduction}
\setcounter{equation}{0} 

The study of integrable partial difference equations (P$\Delta$Es) dates back to the pioneering work of Ablowitz and 
Ladik, \cite{AL}, and of Hirota, \cite{Hir}, motivated partly by the search for numerical finite-difference 
schemes that are discrete in time as well as in space. 
Such systems constitute discrete analogues of soliton type PDEs.   
In subsequent work the Lie-algebraic approach of the Kyoto school, \cite{DJM}, on the one hand, and the approach using 
a so-called \textit{direct linearisation} method, cf. \cite{NQC,QNCL}, on the other hand led to new systematic 
constructions for such systems. In recent years they have been reinvestigated from various points of view, including 
reductions to integrable dynamical mappings \cite{PNC,QCPN}, and associated finite-gap solutions \cite{BP,EN,MP}, 
reductions to discrete Painlev\'e equations, \cite{NP,NRGO} and similarity reductions, whilst soliton solutions arose as a direct 
corollary from the original constructions mentioned above.   

The property of multidimensional consistency, first set out explicitly in \cite{NW}, cf. also \cite{BS}, lies implicitly within 
lattice equations which have the interpretation of a superposition principle for B\"acklund transformations. 
This was identified in \cite{NRGO} as the property constituting the precise discrete analogue of the existence of hierarchies of 
nonlinear evolution equations, and hence of integrability. 
The property has subsequently been used by Adler, Bobenko and Suris \cite{ABS,ABS2} as a 
classifying property. Within certain additional conditions they produced a full list of scalar
quadrilateral lattice equations which are multidimensionally consistent. This list, which surprisingly is quite 
short, is reminiscent of Painlev\'e's list of transcendental equations in the case  
of second order ODEs possessing the property of non-movable singularities of the general solution. We reproduce this 
list from \cite{ABS} below\footnote{Note that the Q4 equation as given by \cite{ABS}, which was first found by Adler 
in \cite{Adler}, is different from the one in \eqref{eq:Qeqsd}, which in this form was first presented in 
\cite{Hie}.} containing three groups of equations: the Q-list the H-list and the A-list: 
\vspace{.2cm}

\noindent
\framebox[2cm][l]{\bf Q-list:} 
\bse\label{eq:Qeqs}\begin{eqnarray}
\fl &&{\rm Q1}: \quad  \po(u-\wh{u})(\wt{u}-\wh{\wt{u}})-\qo(u-\wt{u})(\wh{u}-\wh{\wt{u}})=\dd^2\po\qo(\qo-\po) 
\label{eq:Qeqsa} \\ 
\fl &&{\rm Q2}: \quad  \po(u-\wh{u})(\wt{u}-\wh{\wt{u}})-\qo(u-\wt{u})(\wh{u}-\wh{\wt{u}})+
\po\qo(\po-\qo)(u+\wt{u}+\wh{u}+\wh{\wt{u}})=\nn \\ 
\fl && \hspace{3cm}  =\po\qo(\po-\qo)(\po^2-\po\qo+\qo^2) \label{eq:Qeqsb}\\ 
\fl &&{\rm Q3}: \quad \po(1-\qo^2)(u\wh{u}+\wt{u}\wh{\wt{u}})-\qo(1-\po^2)(u\wt{u}+\wh{u}\wh{\wt{u}})= \nn \\ 
\fl && \hspace{3cm} =(\po^2-\qo^2)\left((\wh{u}\wt{u}+u\wh{\wt{u}})+\dd^2\frac{(1-\po^2)(1-\qo^2)}{4\po\qo}\right) 
\label{eq:Qeqsc} \\
\fl &&{\rm Q4}: \quad \po(u\wt{u}+\wh{u}\wh{\wt{u}})-\qo(u\wh{u}+\wt{u}\wh{\wt{u}})= \nn \\ 
\fl && \hspace{3cm} =\frac{\po\Qo-\qo\Po}{1-\po^2\qo^2}\left((\wh{u}\wt{u}+u\wh{\wt{u}})-
\po\qo(1+u\wt{u}\wh{u}\wh{\wt{u}})\right) \label{eq:Qeqsd}
\end{eqnarray}\ese 
where ~$\Po^2=\po^4-\gm \po^2+1$~,~$\Qo^2=\qo^4-\gm \qo^2+1$~.
\vspace{.2cm}

\noindent
\framebox[2cm][l]{\bf H-list:} 
\bse\label{eq:Heqs}\begin{eqnarray}
\fl &&{\rm H1}: \quad (u-\wh{\wt{u}})(\wt{u}-\wh{u})=\po-\qo  \label{eq:Heqsa}\\ 
\fl &&{\rm H2}: \quad (u-\wh{\wt{u}})(\wt{u}-\wh{u})=(\po-\qo)(u+\wt{u}+\wh{u}+\wh{\wt{u}})+\po^2-\qo^2  
\label{eq:Heqsb}\\ 
\fl &&{\rm H3}: \quad \po(u\wt{u}+\wh{u}\wh{\wt{u}})-\qo(u\wh{u}+\wt{u}\wh{\wt{u}})=\dd(\qo^2-\po^2) 
\label{eq:Heqsc}
\end{eqnarray}\ese 
\vspace{.2cm}

\noindent
\framebox[2cm][l]{\bf A-list:}  
\bse\label{eq:Aeqs}\begin{eqnarray}
\fl &&{\rm A1}: \quad \po(u+\wh{u})(\wt{u}+\wh{\wt{u}})-\qo(u+\wt{u})(\wh{u}+\wh{\wt{u}})=\dd^2\po\qo(\po^2-\qo^2) \label{eq:Aeqsa}\\ 
\fl &&{\rm A2}: \quad \po(1-\qo^2)(u\wh{u}+\wt{u}\wh{\wt{u}})-\qo(1-\po^2)(u\wt{u}+\wh{u}\wh{\wt{u}}) +(\po^2-\qo^2)\left(1+u\wt{u}\wh{u}\wh{\wt{u}}\right)=0 \label{eq:Aeqsb}
\end{eqnarray}\ese 
\vspace{.2cm}

\begin{figure}[t]
\begin{center}  

\setlength{\unitlength}{0.8mm}
\begin{picture}(40,40)(10,0) 

\put(0,0){\circle*{3}}
\put(0,30){\circle*{3}}
\put(30,30){\circle*{3}}
\put(30,0){\circle*{3}}

\put(0,0){\vector(1,0){15}}
\put(15,0){\vector(1,0){15}}
\put(0,30){\vector(0,-1){15}}
\put(0,15){\vector(0,-1){15}}
\put(0,30){\vector(1,0){15}}
\put(15,30){\vector(1,0){15}}
\put(30,30){\vector(0,-1){15}}
\put(30,15){\vector(0,-1){15}}

\put(-7,-7){$\wh{u}$}
\put(34,-5){$\wh{\wt{u}}$}
\put(34,34){$\wt{u}$}
\put(-5,34){$ u$}

\put(15,-5){$\po$}
\put(15,33){$\po$}
\put(-6,15){$\qo$}
\put(34,15){$\qo$}
\end{picture} 
\end{center} 
\caption{Arrangement of the shifted dependent variable on the vertices of a quadrilateral and association of the lattice parameters to the edges.}
\label{fig:quad}
\end{figure}
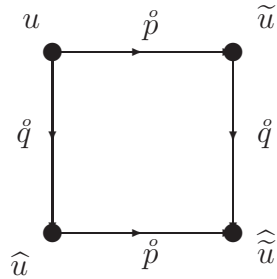

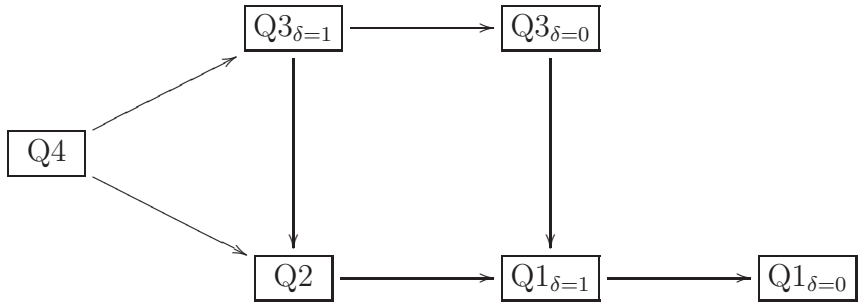
\begin{figure}[h]
\xymatrix{
&&& \framebox{Q3$_{\delta=1}$} \ar[rr] \ar[dd] && \framebox{Q3$_{\delta=0}$} \ar[dd]\\
&\framebox{\ Q4\ } \ar[urr] \ar[drr]\\
&&& \framebox{\ Q2\ } \ar[rr] && \framebox{Q1$_{\delta=1}$} \ar[rr]&& \framebox{Q1$_{\delta=0}$}\\
}
\caption{
\label{fig:Q}
Coalescence diagram for equations in the ABS Q list (this was first given by Adler and Suris in \cite{Q4}).
}
\end{figure}

\noindent
The notation we have adopted here and in earlier papers is the following:  
the vertices along an elementary plaquette on a rectangular lattice contain the 
dependent variables: 
$$ u:=u_{n,m},\qquad \wt{u}=u_{n+1,m}, $$
$$ \wh{u}:=u_{n,m+1}, \qquad \wh{\wt{u}}=u_{n+1,m+1}, $$
which schematically are indicated in Figure \ref{fig:quad},  
in which $\po$ and $\qo$ denote \textit{lattice parameters} associated with the 
directions in the lattice (measuring the grid size in these directions): 
$u~\stackrel{\po}{\rightarrow}~ \wt{u}$,  
$u~\stackrel{\qo}{\rightarrow}~ \wh{u}$. 
The lattice parameters associated to the two directions on the lattice play a central role in the notion of multidimensional consistency.
Specifically they parametrise the family of equations which are compatible on the multidimensional lattice.
The equations in the Q-list are related by degeneration (of the elliptic curve associated to Q4) through the coalescence scheme illustrated in figure \ref{fig:Q}.
The H-list and A-list appear also by degeneration from the Q-list, leading in principle to an extension of Figure \ref{fig:Q}, and we will  
make some of these connections explicit in the later part of the present paper. 
Connections between the ABS equations are not limited to coalescence by degeneration, there also exist Miura and B\"acklund type transformations connecting distinct equations in the list, some transformations of B\"acklund type were discussed recently in \cite{Atkinson}.

The present paper is part I of a sequence of papers dedicated to closed-form $N$-soliton 
solutions of the lattice equations mentioned above. Part II, by Hietarinta and Zhang, \cite{HieZhang}, will be 
dedicated to the Casorati form of the same solutions, establishing a different methodology. 
The results in the present paper are based on a Cauchy matrix structure which goes back to earlier work \cite{NQC,QNCL}, on the so-called 
\textit{direct linearization approach}. In section 2 we review the application of that approach to soliton solutions of lattice equations 
``of KdV type'', which were cases already known in the earlier papers mentioned. Specifically these include the lattice potential KdV, potential 
modified KdV, Schwarzian KdV and an interpolating equation between them which has been referred to as the NQC equation (cf \cite{RasinNQC}). 
(These equations are equivalent to the equations H1, H3$_{\delta=0}$, Q1$_{\delta=0}$ and Q3$_{\delta=0}$ respectively up to gauge transformations.)
The machinery introduced in section 2 will comprise 
several objects, and relations between them, which we will need throughout the remainder of the paper, because the 
$N$-soliton solution for the full Q3 equation can be expressed in terms of those quantities. A constructive 
proof of this $N$-soliton solution is presented in section 3, whilst in section 4 we show that this $N$-soliton solution is consistent with the 
B\"acklund transformation (i.e., defining the latter by a copy of the lattice equation itself we establish the precise relation between 
the $N$- and $N+1$-soliton solution in terms of this BT).  As an upshot of the present treatment, it becomes apparent that the natural parametrisation 
of Q3 turns out to involve already an elliptic curve, whose branch points can be viewed as lattice parameters associated with additional lattice directions.
Degeneration from Q3 in this parametrisation by bringing together one or more branch points of the curve yields one or other of the equations on the ABS list.
In section 5 the degenerations from this parametrisation of Q3 are given in detail and we construct the $N$-soliton solutions for all the equations in the 
ABS list, except for the elliptic case of Q4. This reveals, on the level of the $N$-soliton solutions, a deep connection between the equations 
expressed in terms of the basic quantities introduced in section 2. Section 6 contains a discussion of the results and concluding remarks.

\section{$N$-Soliton solutions of KdV Type Lattice equations}
\setcounter{equation}{0} 

Among the equations in the ABS list we distinguish a subclass which we call \textit{of KdV type}, and they comprise the 
lattice equations which have appeared already many years ago in the context of discretzations of the KdV equations and 
its counterparts, cf e.g. \cite{Hir,NQC,QNCL}. 
In this section we introduce objects from which we may construct solutions of the KdV type lattice equations.

\subsection{Cauchy matrix and recurrence structure}  

We will start by introducing the following Cauchy type matrix: 
\be\label{eq:Cauchymat} 
\bM=\left(M_{i,j}\right)_{i,j=1,\dots,N}\quad ,\quad  M_{i,j}\equiv \frac{\rho_ic_j}{k_i+k_j}\  , \ee  
which will form the core of the structure which we will develop. 
In \eqref{eq:Cauchymat} the $c_i$, $k_i$, ($i=1,\dots,N$) denote two sets of $N$ nonvanishing parameters, 
which we may chose freely (apart that we have to assume that $k_i+k_j\neq 0$, $\forall i,j=1,\dots,N$, 
in order to avoid difficulties with the numerators in the matrix $\bM$). These parameters are 
assumed not to depend on the lattice variables, i.e. on the discrete independent variables $n$ and $m$. The dependence on 
the latter are incorporated wholly in the functions $\rho_i$, the so-called \textit{plane wave factors}, which have the form: 
\be\label{eq:rho} 
\rho_i= \left(\frac{p+k_i}{p-k_i}\right)^n \left(\frac{q+k_i}{q-k_i}\right)^m \rho_i^0  \ , 
\ee 
where the $\rho_i^0$ are initial values, i.e. constant wih regard to the variables $n$,$m$\footnote{Importantly, the $\rho_i^0$ can still 
contain an in principle arbitrary number of additional discrete exponential factors of the form given in \eqref{eq:rho}, i.e. depending on 
additional lattice variables, each of which would be associated with its own lattice parameter. Thus, a more general form for the $\rho_i$ would be 
$$\rho_i= \prod_{\nu\atop p_\nu\neq \pm k_i}\left(\frac{p_\nu+k_i}{p_\nu-k_i}\right)^{n_\nu}\   ,  $$ 
containing an arbitrary number of lattice variables $n_\nu$ each with lattice parameter $p_\nu$ labelled by some index 
$\nu$. Every statement 
derived below can be extended without restriction to involve any choice of these variables, and in particular this implies the multidimensional 
consistency of the equations derived from the scheme. Thus, we will say that these results can be \textit{covariantly extended} to the multidimensional case.}.  

Let us now introduce for convenience the following notation. Let $\bK$ denote the diagonal $N\times N$ matrix containing 
the parameters $k_i$ on the diagonal, and introduce a column vector $\brr$, containing the entries $\rho_i$, and 
a row vector $\tbc$, containing the entries $c_i$, i.e. 
\be\label{eq:Krc}
\bK=\left(\begin{array}{cccc} k_1 & & & \\ 
 & k_2 & & \\ 
& & \ddots & \\ 
& & & k_N \end{array}\right) \quad,\quad \brr=\left(\begin{array}{c}\rho_1\\ \rho_2 \\ \vdots \\ \rho_N\end{array}\right) 
\quad,\quad \tbc=\left( c_1,c_2,\cdots,c_N\right)\  . 
\ee

It is easily checked that we have from the definition \eqref{eq:Cauchymat} immediately the relation:  
\be\label{eq:dyadic1}
\bM\,\bK+\bK\,\bM=\brr\,\tbc\  , 
\ee 
where significantly the \textit{dyadic} on the right hand side is a matrix of rank 1. 

Next we establish the dynamics in terms of  the matrix $\bM$, which follows from the definition \eqref{eq:Cauchymat} together 
with the dynamical equations for $\rho_i$ \eqref{eq:rho} which are simply 
\be\label{eq:rhoshift}
\wt{\rho}_i=T_p\rho_i=\rho_i(n+1,m)=\frac{p+k_i}{p-k_i}\,\rho_i\quad,\quad  
\wh{\rho}_i=T_q\rho_i=\rho_i(n,m+1)=\frac{q+k_i}{q-k_i}\,\rho_i\  ,  
\ee  
where  $T_p$, $T_q$ denote the elementary shift operators in the lattice in the directions associated with 
lattice parameters $p$ and $q$ respectively, $T_{-p}=T_p^{-1}$ and $T_{-q}=T_q^{-1}$ denoting their inverses. 
A straightforward calculation then yields the following relations 
\bse\label{eq:Mrels}  \begin{eqnarray}
&& \wt{\bM}\,(p\bun+\bK)-(p\bun+\bK)\,\bM =\wt{\brr}\,\tbc\  , \label{eq:Mrelsa} \\ 
&&\wh{\bM}\,(q\bun+\bK)-(q\bun+\bK)\,\bM =\wh{\brr}\,\tbc\  , \label{eq:Mrelsb}  
\end{eqnarray}
in which $\bun$ is the $N\times N$ unit matrix, and where we used the obvious notation that the shifts ~$\wt{\phantom{a}}$~, ~$\wh{\phantom{a}}$~
act on all the relevant objects depending on $n,m$ by the respective shifts by one unit in these independent variables. In addition, we have the adjoint 
relations
\begin{eqnarray}
&& (p\bun-\bK)\,\wt{\bM}-\bM\,(p\bun-\bK)=\brr\,\tbc\  , \label{eq:Mrelsc} \\ 
&& (q\bun-\bK)\,\wh{\bM}-\bM\,(q\bun-\bK)=\brr\,\tbc\ . \label{eq:Mrelsd}
\end{eqnarray}\ese 
The equations \eqref{eq:Mrels} encode all the information on the dynamics of the matrix $\bM$, w.r.t. the discrete variables $n$, $m$, in addition to 
\eqref{eq:dyadic1} which can be thought of as the defining property of $\bM$.  

Now we introduce several objects involving the matrix $\bM$, in terms of which we can define the basic variables 
which will solve the relevant lattice equations. Thus, we introduce the determinant 
\begin{equation}\label{eq:taudet}
f=f_{n,m}=\det\left(\boldsymbol{1}+\boldsymbol{M}\right) \  , 
\end{equation} 
which we will identify later as the relevant $\tau$-function (obeying Hirota-type bilinear equations), 
as well as the following quantities\footnote{Here and earlier we use the symbols $\tbc$ and $\tbu$ to denote 
\textit{adjoint} row vectors, which in the case of the first is just the transposed of the column vector $\bc$, i.e. 
$\tbc=\bc^T$, but in the case of $\tbu$ is not simply the transposed of the column vector $\bu$, but rather a quantity 
that is defined in \eqref{eq:tu} in its own right. Thus, the left super-index ``t'' should no be confused with the  operation of transposition, but rather indicate a new object obeying some linear equations associated with the vector 
$\bu$.} 
\bse\label{eq:defsU}\bea
\bu^{(i)} &=& \left( \bun+\bM\right)^{-1}\,\bK^i\,\brr    \label{eq:u} \\ 
\tbu^{(j)} &=& \tbc\,\bK^j\,\left( \bun+\bM\right)^{-1}    \label{eq:tu} \\ 
S^{(i,j)} &=& \tbc\,\bK^j\,\left( \bun+\bM\right)^{-1}\,\bK^i\,\brr\  ,     \label{eq:U} 
\eea\ese 
for $i,j\in\mathbb{Z}$ (assuming that none of the parameters $k_i$ is zero).  
Thus, we obtain an infinite sequence of column vectors $\bu^{(i)}$, of row vectors $\tbu^{(j)}$ and a 
infinte by infinite array of scalar quantities $S^{(i,j)}$. An important property of the latter objects, 
which can also be written as 
\be\label{eq:UU} 
 S^{(i,j)}= \tbc\,\bK^j\,\bu^{(i)}=\tbu^{(j)}\,\bK^i\,\brr\  , \ee  
is that they are symmetric w.r.t. the interchange of the indices, i.e. 
\be\label{eq:Usymm} S^{(i,j)}=S^{(j,i)}, \ee 
provided that the constants $c_i$, $\rho_i^0$, $k=1\dots N$, are all nonzero.  

We shall now derive, starting from \eqref{eq:Mrels} a system of recurrence relations 
which describe the dynamics for the quantities defined in \eqref{eq:defsU}. Once this recursive structure is 
established, we will single out specific combinations of the $S^{(i,j)}$ in terms of which we can derive closed form 
discrete equations. 
In fact, from the definition \eqref{eq:u}, using the fact that 
$\bK$ is a diagonal matrix, we have: 
\begin{eqnarray*}
&& \bK^i\,\brr = (\bun+\bM)\,\bu^{(i)}\quad \Rightarrow\quad 
\bK^i\,\wt{\brr} = (\bun+\wt{\bM})\,\wt{\bu}^{(i)}\quad  \\ 
&& \Rightarrow\quad \bK^i\,\frac{p\bun+\bK}{p\bun-\bK}\,\brr= (\bun+\wt{\bM})\,\wt{\bu}^{(i)} \\ 
&& \Rightarrow\quad \bK^i\,(p\bun+\bK)\,\brr= (p\bun-\bK)\,(\bun+\wt{\bM})\,\wt{\bu}^{(i)} \\ 
&& \qquad = (p\bun-\bK)\,\wt{\bu}^{(i)} +(p\bun-\bK)\,\wt{\bM}\,\wt{\bu}^{(i)} \\ 
&& \qquad =  (p\bun-\bK)\,\wt{\bu}^{(i)} + \bM\,(p\bun-\bK)\,\wt{\bu}^{(i)}+ \brr\,\tbc \wt{\bu}^{(i)} \\ 
\end{eqnarray*}
where in the last step use has been made of \eqref{eq:Mrelsc}. Using now \eqref{eq:U} we 
we conclude that
$$  p\bK^j\,\brr+\bK^{j+1}\,\brr=(\bun+\bM)\,(p\bun-\bK)\,\wt{\bu}^{(i)}+ \wt{S}^{(i,0)}\brr\  . $$ 
Multiplying both sides by the inverse matrix ~$(\bun+\bM)^{-1}$~ and identifying the terms on the 
left hand side using (\eqref{eq:u}, we thus obtain  
\begin{eqnarray} 
(p\bun-\bK)\,\wt{\bu}^{(i)}&=&(\bun+\bM)^{-1}\,\left[ p\bK^i\,\brr+\bK^{i+1}\,\brr-\wt{S}^{(i,0)}\brr \right] \nn \\ 
&=& p\bu^{(i)}+\bu^{(i+1)}-\wt{S}^{(i,0)}\bu^{(0)}\  . \label{eq:recurs1} 
\end{eqnarray} 
Thus, we have obtained a linear recursion relation between the objects $\bu^{(i)}$ with the objects 
$S^{(i,j)}$ acting as coefficients. In quite a similar fashion we can derive the relation 
\be\label{eq:recurs2}  
(p\bun+\bK)\,\bu^{(i)} = p\wt{\bu}^{(i)}-\wt{\bu}^{(i+1)}+S^{(i,0)}\wt{\bu}^{(0)}\  , \ee 
in fact by making use of \eqref{eq:Mrelsb} in this case. Eq. \eqref{eq:recurs2} can be thought of as 
an inverse relation to \eqref{eq:recurs1}, noting that the $\wt{\phantom{a}}$-shifted objects are now at the 
right hand side of the equation. Multiplying both sides of either \eqref{eq:recurs1} or \eqref{eq:recurs2} 
from the left by the row vector ~$\tbc\,\bK^j$~ and identifying the resulting terms through \eqref{eq:U}, 
we obtain now a relation purely in terms of the objects $S^{(i,j)}$, namely: 
\begin{eqnarray}  
&& \tbc\,\bK^j\,(p\bun-\bK)\,\wt{\bu}^{(i)} = \tbc\,\bK^j\,\left[p\bu^{(i)}+\bu^{(i+1)}-\wt{S}^{(i,0)}\bu^{(0)}\right] \nn \\ 
&&\qquad \Rightarrow\quad  p\wt{S}^{(i,j)}-\wt{S}^{(i,j+1)}=pS^{(i,j)}+S^{(i+1,j)}-\wt{S}^{(i,0)} S^{(0,j)} \label{eq:Recurs1} 
\end{eqnarray}  
using the fact that $\wt{S}^{(i,0)}$ is just a scalar factor which can be moved to the left of the matrix multiplication. 
Thus, we have now obtained a \textit{nonlinear} recursion relation between the $S^{(i,j)}$ and its 
~$\wt{\phantom{a}}$-shifted counterparts. 

In a similar fashion, multiplying eq. \eqref{eq:recurs2} by the row vector ~$\tbc\,\bK^j$~ we obtain the 
complementary relation: 
\be\label{eq:Recurs2} 
pS^{(i,j)}+S^{(i,j+1)}= p \wt{S}^{(i,j)}-\wt{S}^{(i+1,j)}+S^{(i,0)}\wt{S}^{(0,j)}\ , 
\ee 
however this relation can be obtained from the previous relation \eqref{eq:Recurs1} by interchanging the indices, using the 
symmetry \eqref{eq:Usymm}. 

\paragraph{Remark:} Combining both eqs. \eqref{eq:recurs1} and \eqref{eq:recurs2}, using also \eqref{eq:Recurs2}, 
the following \textit{algebraic} (i.e. not involving lattice shifts) recurrence relation can be derived: 
\be\label{eq:algrel}
\bK^2\bu^{(i)}=\bu^{(i+2)}+ S^{(i,1)}\,\bu^{(0)}-S^{(i,0)}\,\bu^{(1)}\  , 
\ee 
which in turn gives rise to the following algebraic recurrence for the objects $S^{(i,j)}$, namely 
\be\label{eq:Algrel}
S^{(i,j+2)}=S^{(i+2,j)}+ S^{(i,1)}\,S^{(0,j)}-S^{(i,0)}\,S^{(1,j)}\  . 
\ee 

\paragraph{}  
To summarise the structure obtained, we note that starting from the Cauchy matrix $\bM$ defined in \eqref{eq:Cauchymat}, 
depending dynamically on the lattice variables through the plane-wave factors $\rho_i$ given in \eqref{eq:rho}, we have 
defined an infinite set of objects, namely column and row vectors $\bu^{(i)}$ and $\tbu^{(i)}$, and a doubly infinite sequence of scalar 
functions $S^{(i,j)}$, all related through a system of dynamical (since it involves lattice shift) recurrence 
relations. Obviously, all relations that we have derived for the ~$\wt{\phantom{a}}$-shifts (involving lattice 
parameter $p$ and lattice variable $n$) hold also for the ~$\wh{\phantom{a}}$-shifts, simply by replacing 
$p$ by $q$ and interchanging the roles of $n$ and $m$. Thus, for the scalar objects $S^{(i,j)}$ we have the following 
set of coupled recurrence relations: 
\bse\label{eq:recurs}\begin{eqnarray}
p\wt{S}^{(i,j)}-\wt{S}^{(i,j+1)}&=& p S^{(i,j)}+S^{(i+1,j)}-\wt{S}^{(i,0)}S^{(0,j)}\ ,  \label{eq:recursa}\\ 
pS^{(i,j)}+S^{(i,j+1)}&=& p \wt{S}^{(i,j)}-\wt{S}^{(i+1,j)}+S^{(i,0)}\wt{S}^{(0,j)}\ ,  \label{eq:recursb}\\ 
q\wh{S}^{(i,j)}-\wh{S}^{(i,j+1)}&=& q S^{(i,j)}+S^{(i+1,j)}-\wh{S}^{(i,0)}S^{(0,j)}\ ,  \label{eq:recursc}\\ 
qS^{(i,j)}+S^{(i,j+1)}&=& q \wh{S}^{(i,j)}-\wh{S}^{(i+1,j)}+S^{(i,0)}\wh{S}^{(0,j)}\ .  \label{eq:recursd} 
\end{eqnarray}\ese 
We proceed now by deriving closed-form lattice equations for individual elements chosen from the $S^{(i,j)}$ as functions of the variables 
$n$, $m$.

\subsection{Closed Form Lattice Equations}

We start with the variable $S^{(0,0)}$, for which we can derive a partial difference equation as follows. In fact, subtracting 
\eqref{eq:recursc} from \eqref{eq:recursa} we obtain 
\be\label{eq:pUqU} 
p\wt{S}^{(i,j)}-q\wh{S}^{(i,j)}-\wt{S}^{(i,j+1)}+\wh{S}^{(i,j+1)}= (p-q) S^{(i,j)}-(\wt{S}^{(i,0)}-\wh{S}^{(i,0)}) S^{(0.j)}\ . \ee 
On the other hand, taking the $\wh{\phantom{a}}$-shift of \eqref{eq:recursb}, let us refer to it as $\wh{\eqref{eq:recursb}}$, and subtracting 
from it the $\wt{\phantom{a}}$-shift of \eqref{eq:recursd}, i.e. $\wt{\eqref{eq:recursd}}$, we obtain 
\be\label{eq:qUpU} 
p\wh{S}^{(i,j)}-q\wt{S}^{(i,j)}+\wh{S}^{(i,j+1)}-\wt{S}^{(i,j+1)}= (p-q) \wh{\wt{S}}_{(i,j)}+(\wh{S}^{(i,0)}-\wt{S}^{(i,0)}) \wh{\wt{S}}^{(0.j)}\ . \ee 
Combining both equations, the terms which have a shift in their second index drop out and we obtain the equation:
\be\label{eq:UpqU} 
(p+q)(\wt{S}^{(i,j)}-\wh{S}^{(i,j)})=(p-q)(S^{(i,j)}-\wh{\wt{S}}^{(i,j)})+(\wh{S}^{(i,0)}-\wt{S}^{(i,0)}) (S^{(0,j)}-\wh{\wt{S}}^{(0,j)})\  . \ee 
Setting now $i=j=0$ in the last formula, we see that we get a closed form equations in terms of $w\equiv S^{(0,0)}$. This yields, after some trivial 
algebra the equation: 
\be\label{eq:dKdV}
(p+q+w-\wh{\wt{w}})(p-q+\wh{w}-\wt{w})=p^2-q^2\  ,  
\ee 
which is the lattice potential KdV equation, which has appeared in the literature in various guises, cf. \cite{Hir,NQC,QNCL}, notably 
as the permutability condition of the B\"acklund transformations for the KdV equation, cf. \cite{WE}.  
Curiously, this integrable partial difference equation can  also be traced back to numerical analysis, where it has appeared 
in the form of the $\epsilon$-algorithm of Wynn, \cite{Wynn}, as an efficient convergence accelerator algorithm. 
In the present context of the structures arising from the Cauchy matrix we have established here an infinite family of solutions of the form  
\be\label{eq:wsol} w\equiv S^{(0,0)}=\tbc\,(\bun+\bM)^{-1}\,\brr \ee  
which constitute the $N$-soliton solutions for the equation \eqref{eq:dKdV}. 

Eq. \eqref{eq:dKdV} is by no means the only equation that emerges from the set of relations \eqref{eq:recurs}. In fact, instead of singling 
out $w=S^{(0,0)}$, we can chose other elements among the $S^{(i,j)}$, 
or (linear) combinations of them, and then systematically investigate what equations these choices satisfy by exploring 
the system of recurrence relations \eqref{eq:recurs}. For example, from \eqref{eq:pUqU} taking $i=0$, $j=-1$ and 
introducing the variable ~$v\equiv 1-S^{(0,-1)}$~ it is a simple exercise to obtain the following relation: 
\be\label{eq:dMiura1}
p-q+\wh{w}-\wt{w}=\frac{p\wt{v}-q\wh{v}}{v}\  . 
\ee 
Alternatively, adding the $\wh{\phantom{a}}$-shift of \eqref{eq:recursa} to \eqref{eq:recursd} we get 
\be\label{eq:UqpU} 
p\wh{\wt{S}}^{(i,j)}+qS^{(i,j)}-\wh{\wt{S}}^{(i,j+1)}+S^{(i,j+1)}= (p+q) \wh{S}^{(i,j)}+(S^{(i,0)}-\wh{\wt{S}}^{(i,0)}) \wh{S}^{(0.j)}\ , \ee 
and when taking in \eqref{eq:UqpU} $i=0$, $j=-1$ an easy calculation yields: 
\be\label{eq:dMiura2}
p+q+w-\wh{\wt{w}}=\frac{p\wh{\wt{v}}+qv}{\wh{v}}\  . 
\ee 
Clearly, in \eqref{eq:dMiura2}, interchanging $p$ and $q$ and the $\wt{\phantom{a}}$-shift and the $\wh{\phantom{a}}$-shift should not make a difference, since the 
left-hand side is invariant under this change. Thus, the right-hand side must be invariant as well, leading to the relation: 
\be\label{eq:dMKdV} 
p\left(v\wh{v}-\wt{v}\wh{\wt{v}}\right)=q\left(v\wt{v}-\wh{v}\wh{\wt{v}}\right)\  .   
\ee 
Eq. \eqref{eq:dMKdV} is an integrable P$\Delta$E in its own right for the quantity $v$, for which, by construction, we have an infinite family of solutions, 
namely given by 
\be\label{eq:MKdVsol} v=1-S^{(0,-1)}=1-\tbc\,\bK^{-1}\,(\bun+\bM)^{-1}\,\brr\  .  \ee 
The P$\Delta$E \eqref{eq:dMKdV} for the variable $v$ is identified as the lattice potential MKdV equation, which also occurred in  
\cite{NQC,QNCL}, and is actually closely related to the lattice sine-Gordon equation of \cite{Hir}. The relations \eqref{eq:dMiura1} and 
\eqref{eq:dMiura2} constitute a Miura transform between the lattice potential MKdV \eqref{eq:dMKdV} and the he lattice potential KdV equation 
\eqref{eq:dKdV}.  

As another choice of dependent variables we can consider is the variable $\UD{-1}{-1}$, i.e. we can consider \eqref{eq:recurs} for $i=j=-1$, leading to
$$ p\left(\wt{S}^{(-1,-1)}+S^{(-1,-1)}\right)=1-\left(1-\wt{S}^{(-1,0)}\right)\,\left( 1-S^{(0,-1)}\right)\  , $$ 
and a similar relation for $p$ replaced by $q$ and the $\wt{\phantom{a}}$-shift replaced by the $\wh{\phantom{a}}$-shift. Using the fact that 
$S^{(-1,0)}=S^{(0,-1)}=1-v$ and introducing the abbreviation ~$z=S^{(-1,-1)}-\frac{n}{p}-\frac{m}{q}$~, the latter relations reduce to: 
\be\label{eq:zv} 
p(z-\wt{z})=\wt{v}v \quad ,\quad q(z-\wh{z})=\wh{v}v \quad . 
\ee 
On the one hand, these two equations lead back to the equation \eqref{eq:dMKdV} by eliminating the variable $z$ (considering in 
addition the $\wt{\phantom{a}}$- and $\wh{\phantom{a}}$-shifts of the two relations). On the other hand, by eliminating the variable $v$ we obtain yet again 
a P$\Delta$E, but now for $z$ which reads
\be\label{eq:dSKdV} 
\frac{(z-\wt{z})(\wh{z}-\wh{\wt{z}})}{(z-\wh{z})(\wt{z}-\wh{\wt{z}})}=\frac{q^2}{p^2}\   , 
\ee 
which we identify with the Schwarzian lattice KdV equation, and is also referred to as \textit{cross-ratio equation}\footnote{Eq. \eqref{eq:dSKdV} was 
first established as an integrable lattice equation in \cite{NC}, but was also studied in connection with discrete conformal function theory, \cite{BP}. 
Interestingly, also this equation has appeared in the context of numerical analysis, in conection with the Pad\'e tables in work by R. Cordellier, \cite{Cordellier}.}. 
Here we have constructed $N$-soliton solutions of eq. \eqref{eq:dSKdV} given explicitly by the formula
\be\label{eq:zsol} z=\tbc\,\bK^{-1}\,(\bun+\bM)^{-1}\,\bK^{-1}\,\brr - z_0 - \frac{n}{p} - \frac{m}{q},  \ee 
in which $z_0$ is an arbitrary constant. 

To summarise, we conclude that through the recurrence structure encoded in eqs. \eqref{eq:recurs} we obtain solutions of various different 
P$\Delta$Es in one stroke. 
We will now proceed further by establishing a large parameter class of additional lattice equations, which also provide us information on the 
bilinear structure of the lattice systems.  

\subsection{Bilinear Equations and the NQC Equation} 
 
We will start by considering the  $\tau$-function for the soliton solutions, given by 
\eqref{eq:taudet}, using the relation \eqref{eq:Mrelsa}, 
we can perform the following straightforward calculation:
\begin{eqnarray*} 
\wt{f} &=& \det\left( \bun+\wt{\bM}\right)=\det\left\{ \bun+\left[ (p\bun+\bK) \bM +\wt{\brr}\tbc\right] (p\bun+\bK)^{-1} \right\} \\  
&=& \det\left\{ (p\bun+\bK) \left[ \bun + \bM + (p\bun+\bK)^{-1}\wt{\brr} \tbc \right] (p\bun+\bK)^{-1} \right\} \\ 
&=& \det\left\{ (\bun + \bM) \left[ \bun + (\bun+ \bM)^{-1} (p\bun+\bK)^{-1}\wt{\brr} \tbc \right]  \right\} \\ 
&=& f\,\det\left\{ \bun + (\bun+ \bM)^{-1} (p\bun+\bK)^{-1}\wt{\brr} \tbc  \right\} 
\end{eqnarray*} 
from which, using also 
$$ \wt{\brr}=\frac{p\bun+\bK}{p\bun-\bK}\,\brr\  , $$ 
 we have 
\be\label{eq:wttau} 
\frac{\wt{f}}{f} = 1 + \tbc\,(\bun+\bM)^{-1}\,(p\bun-\bK)^{-1}\,\brr =1+\tbu^{(0)}\,(p\bun-\bK)^{-1}\,\brr\   ,  
\ee
where in the last step we have made use of the determinant relation (a special case of the famous Weinstein-Aronszajn formula):   
$$  \det\left( \bun+\mbx\,\mby^T\right) = 1+\mby^T\cdot\mbx\  ,  $$ 
for arbitrary $N$-component vectors $\mbx$, $\mby$ (the suffix $T$ denoting transposition). 

The combination emerging on the right-hand side of  eq. \eqref{eq:wttau} is a new object which we need in the scheme, and it is 
natural to try and derive some equations for it along the lines of the derivations in subsection 2.1.  
Thus, more generally, let us introduce the function: 
\be\label{eq:Vdef}  
V(a)\equiv 1-\tbc\,(a+K)^{-1}\bu^{(0)}= 1-\tbu^{(0)}\,(a\bun+\bK)^{-1}\,\brr \   , 
\ee 
for any value of a parameter $a\in\mathbb{C}$. From \eqref{eq:wttau} we immediately have 
\be\label{eq:Vtau}
\frac{\wt{f}}{f}=\frac{T_pf}{f}=V(-p)\   ,  
\ee 
but we need further relations involving $V(a)$ to derive closed-form equations for the
$\tau$-function $f$. To do that we introduce some further objects, namely:
\bse\label{eq:adefs}\begin{eqnarray}
\bu(a) &=& (\bun+\bM)^{-1}(a\bun+\bK)^{-1}\brr \  , \label{eq:adefsa} \\ 
\tbu(b) &=& \tbc\,(b\bun+\bK)^{-1}(\bun+\bM)^{-1} \  , \label{eq:adefsb} \\ 
S(a,b) &=& \tbc\,(b\bun+\bK)^{-1}(\bun+\bM)^{-1}(a\bun+\bK)^{-1}\brr \  . \label{eq:adefsc} 
\end{eqnarray}\ese 
It can be shown that the latter object has the symmetry: 
\be\label{eq:Uabs}
S(a,b)=\tbc\,(b\bun+\bK)^{-1}\bu(a)=\tbu(b)\,(a\bun+\bK)^{-1}\brr=S(b,a)\  , 
\ee 
in which $a,b\in \mathbb C$ are arbitrary parameters. 

Following a similar derivation as the one leading to \eqref{eq:recurs1} 
and \eqref{eq:recurs2}, derive the following relations:
\bse\label{eq:VRecurs}\begin{eqnarray} 
(p\bun-\bK)\wt{\bu}(a)&=& \wt{V}(a)\bu^{(0)}+(p-a)\bu(a)\  ,  \label{eq:VRecurs1} \\ 
(p\bun+\bK)\bu(a)&=& -V(a)\wt{\bu}^{(0)}+(p+a)\wt{\bu}(a)\  ,  \label{eq:VRecurs2} 
\end{eqnarray}\ese 
Furthermore, by multiplying \eqref{eq:VRecurs} from the left by the row vector 
~$\tbc\,(b\bun+\bK)^{-1}$~, show that \eqref{eq:VRecurs} leads to 
\bse\label{eq:Uab}\be \label{eq:Uaba}
1-(p+b)\,\wt{S}(a,b)+(p-a)\,S(a,b) =  \wt{V}(a)\,V(b)\   . 
\ee 
This and its companion equation 
\be \label{eq:Uabb}
1-(q+b)\,\wh{S}(a,b)+(q-a)\,S(a,b) =  \wh{V}(a)\,V(b)\   . 
\ee \ese 
form one of the basic relations for the rest of this paper. In fact, by setting $a=-p, b=-q$ in 
\eqref{eq:Uabb} we obtain 
\be\label{eq:Utau}
1+(p+q)S(-p,-q)= \frac{T_pT_qf}{f}=\frac{\wh{\wt{f}}}{f}\   , 
\ee 
which allows us to express the quantity $S(p,q)$ in terms of the $\tau$-function. 

We will now now use both relations \eqref{eq:Uaba} and \eqref{eq:Uabb}, together with the  the symmetry 
~$S(a,b)=S(b,a)$~, cf. \eqref{eq:Uabs}, to deduce now a partial difference equation for $S(a,b)$ for any fixed $a$, $b$. In fact, 
from the identity:
$$ 
\frac{\wh{\wt{V}(a)\,V(b)}}{\wt{\wh{V}(a)\,V(b)}}=
\frac{\wh{V}(b)\,V(a)}{\wt{V}(b)\,V(a)}
$$ 
by inserting \eqref{eq:Uab} and its counterpart, with $p$ replaced by $q$ and the $\wt{\phantom{a}}$-shift replaced by the 
$\wh{\phantom{a}}$-shift, as well as the relations with $a$ and $b$ interchanged, we obtain the following closed form equation 
for $S(a,b)$: 
\be\label{eq:Uabeq}
\frac{1-(p+b)\,\wh{\wt{S}}(a,b)+(p-a)\,\wh{S}(a,b)}{1-(q+b)\,\wh{\wt{S}}(a,b)+(q-a)\,\wt{S}(a,b)}  
=\frac{1-(q+a)\,\wt{S}(a,b)+(q-b)\,S(a,b)}{1-(p+a)\,\wh{S}(a,b)+(p-b)\,S(a,b)}\   .  
\ee 
In eq. \eqref{eq:Uabeq} the parameters $a$ and $b$ are assumed fixed, and for each choice of them we have 
a quadrilateral P$\Delta$E which is integrable in the sense of the multidimensional consistency property,  
explained in e.g. refs. \cite{NW,BS}, where $p$ and $q$ play the role of lattice parameters. We will elucidate 
the role of the parameters $a$, $b$ in subsequent sections. It suffices here to observe that by fixing the special 
choice $a=p$, $b=-p$ or $a=q$, $b=-q$ respectively in \eqref{eq:Uab}, 
we obtain the relations  
\be\label{eq:VV}
\wt{V}(p)V(-p)=1\quad ,\quad  
\wh{V}(q)V(-q)=1\   . 
\ee 
There are additional relations, connecting the object $V(a)$ with the variable $w$ defined in \eqref{eq:wsol}, 
which will play an important role in the proofs of the main results. Such relations can be obtained by 
eliminating the terms containing the product $\bK\bu(a)$ in the relations \eqref{eq:VRecurs1} and 
\eqref{eq:VRecurs2} and their counterparts in the other lattice direction, and subsequently multiplying the 
the resulting combinations from the left by $\tbc$, using the fact that $\tbc\,\bK\,\bu(a)=1-V(a)$~,~ 
$\tbc\,\bu_0=w$. Thus, we arrive at the following list of relations:  
\bse\label{eq:UVW}\bea
p-q+\wh{w}-\wt{w}&=& (p+a)\frac{\wt{V}(a)}{V(a)}-(q+a)\frac{\wh{V}(a)}{V(a)} \label{eq:UVWa}\\ 
&=& (p-a)\frac{\wh{V}(a)}{\wh{\wt{V}}(a)}-(q-a)\frac{\wt{V}(a)}{\wh{\wt{V}}(a)}\  .   \label{eq:UVWb} \\ 
p+q+w-\wh{\wt{w}}&=& (p+a)\frac{\wh{\wt{V}}(a)}{\wh{V}(a)}+(q-a)\frac{V(a)}{\wh{V}(a)}\label{eq:UVWc} \\ 
&=& (p-a)\frac{V(a)}{\wt{V}(a)}+(q+a)\frac{\wh{\wt{V}}(a)}{\wt{V}(a)}\  .   \label{eq:UVWd} 
\eea \ese 
As a direct corollary the equality on the right-hand sides of eqs. \eqref{eq:UVWa} and \eqref{eq:UVWb}, or 
equivalently \eqref{eq:UVWc} and \eqref{eq:UVWd},  
actually provide us with another integrable lattice equation for the variable $V(a)$. Once again, the parameter $a$ 
plays a distinct role in this equation from the lattice parameters $p$ and $q$, and in terms of the latter 
parameters the equation for $V(a)$ is multidimensionally consistent. 

\paragraph{Remark:} Note that closed-form equations for $V(a)$ are obtained by equating the right-hand sides of \eqref{eq:UVWa} 
and \eqref{eq:UVWb}, or equivalently the right-hand sides of \eqref{eq:UVWc} and \eqref{eq:UVWd}. Furthermore, breaking the 
covariance between the lattice directions by choosing $a=p$ in \eqref{eq:UVW} we get for $V(p)$ the following quadrilateral P$\Delta$E 
\be\label{eq:VPdE}
2p\frac{\wt{V}(p)}{V(p)}=(p+q)\frac{\wh{V}(p)}{V(p)} 
+(p-q)\frac{\wt{V}(p)}{\wh{\wt{V}}(p)}\  .   
\ee 
Although this equation is not multidimensionally consistent in the strong sense (demanding consistency of the same equation 
in all lattice directions) it is multidimensionally consistent in a weaker sense (consistency between different equations 
on different sublattices). In fact, we can supplement \eqref{eq:VPdE} by a similar equation with a lattice variable $h$ instead of $m$, 
and with lattice parameter $q$ replaced by $r$. These two 3-term relation are consistent-around-the-cube with a 4-term lattice equation 
for $V(p)$ of the form arising from the right-hand sides of \eqref{eq:UVWa}, \eqref{eq:UVWb} in the lattice directions associated with
parameters $q$ and $r$.    

\paragraph{} 
Using now \eqref{eq:Vtau} to substitute the variable $V(p)$, and similarly doing the same for 
$V(q)$ in the analogous equation obtained by interchanging $p$ and $q$ and $\wt{\phantom{a}}$-shifts and 
$\wh{\phantom{a}}$-shifts, we obtain the following to \textit{bilinear} partial difference equations for the $\tau$-function $f$, 
namely 
\bse\label{eq:taueq}\begin{eqnarray}
&& (p+q)\ut{\wh{f}}\wt{f}+(p-q)\ut{f}\wh{\wt{f}}=2pf\wh{f} \  , \label{eq:taua} \\ 
&& (p+q)\hypohat 0 {\wt{f}} \wh{f}+(q-p){\hypohat 0 f} \wh{\wt{f}}=2qf\wt{f} \  . \label{eq:taub}
\end{eqnarray}\ese 
Here the under-accents $\underaccent{\wtilde}{f}$ and $\underaccent{\what}{f}$ denote lattice shifts in the opposite 
directions to $\wt{f}$ and $\wh{f}$ respectively. 
It can be shown that these two 6-point equations are consistent on the two-dimensional lattice from an initial-value point of view, but 
we will not go into this here, cf. \cite{NRGO}. We just mention that by 
performing this computation explicitly, and eliminating intermediate values on vertices in the lattice one can derive from \eqref{eq:taueq} 
the following 5-point lattice equation:  
\be\label{eq:bilToda} 
(p-q)^2 \underaccent{\what}{\underaccent{\wtilde}{f}}\wh{\wt{f}}-
(p+q)^2 {\hypohat 0 {\wt{f}}}\,{\hypotilde 0 {\wh{f}}}+4pqf^2=0\  . 
\ee 
which is Hirota's \textit{discrete-time Toda equation}, cf. \cite{Hir}.

\subsection{Lattice KdV and lattice MKdV} 

We finish this section by presenting the actual lattice KdV, which is related to the lattice \textit{potential} KdV equation \eqref{eq:dKdV} 
by considering differences of the variables $w$ along the diagonals. These differences can be expressed in terms of the $\tau$-function, namely 
by setting $a=p$ or $a=-p$ in eqs. \eqref{eq:UVW}, leading to the relations 
\bse\label{eq:UV}\begin{eqnarray}
\Xi\equiv p-q+\wh{w}-\wt{w}&=&(p-q)\frac{\wh{\wt{f}}f}{\wt{f}\wh{f}}\   ,  \label{eq:UVa}\\ 
\Upsilon\equiv p+q+w-\wh{\wt{w}}&=&(p+q)\frac{\wt{f}\wh{f}}{f\wh{\wt{f}}}\   ,  \label{eq:UVb} 
\end{eqnarray}\ese 
from which we have immediately
\be\label{eq:RQ}
\Xi\Upsilon=p^2-q^2\quad,\quad \Xi-\wh{\wt{\Xi}}=\wh{\Upsilon}-\wt{\Upsilon}\  , 
\ee 
which leads to the lattice KdV equation, \cite{Hir}, in terms of either $\Xi$ or $\Upsilon$, by eliminating one or the other variable using the first relation, i.e.  
\begin{equation}\label{eq:DKdV}
\Xi-\wh{\wt{\Xi}}=(p^2-q^2)\left(\frac{1}{\wh{\Xi}}-\frac{1}{\wt{\Xi}}\right)\quad\Leftrightarrow \quad 
\wh{\Upsilon}-\wt{\Upsilon}=(p^2-q^2)\left(\frac{1}{\Upsilon}-\frac{1}{\wh{\wt{\Upsilon}}}\right)\,    . 
\end{equation} 
What is not well known is that the resulting equation admits a \textit{scalar} Lax pair of the form 
\bse\label{eq:KdVlax} \begin{eqnarray}
\wh{\wt{\varphi}} &=& \Upsilon\wh{\varphi}+\ld\varphi\   , \\ 
\wt{\varphi} &=& \wh{\varphi}+\Xi\varphi\   ,   
\end{eqnarray}\ese 
where $\ld=k^2-q^2$ is the spectral parameter.  
In a similar way, from potential lattice MKdV equation \eqref{eq:dMKdV}, by considering ratios of the variable $v$ over the diagonals in the lattice, 
we can obtain the following lattice equation for the variable ~$W\equiv \what{v}/\wt{v}$~
\begin{equation}\label{eq:MKdV}
\frac{\wh{\wt{W}}}{W}=\frac{(p\wh{W}-q)}{(p-q\wh{W})}\,\frac{(p-q\wt{W})}{(p\wt{W}-q)}\   . 
\end{equation}
The P$\Delta$E \eqref{eq:MKdV}, which we identify with the (non-potential) lattice MKdV equation, arises as the compatibility 
condition of the following Lax pair
\bse\begin{eqnarray}
\wh{\psi} &=& W\wt{\psi}+\left(q-pW\right)\psi\  , \\ 
\wh{\wt{\psi}} &=& \frac{p^2-q^2}{p-qW}\,\wh{\psi}+\ld\frac{pW-q}{p-qW}\,\psi\  ,  
\end{eqnarray}\ese
where $\psi$ is a scalar function and in which $\lambda$ is again a spectral parameter. For the variable $W$ we have the following identifications 
in terms of the $\tau$-function: 
\begin{equation}\label{eq:Wtau}
v=V(0)=\frac{T_0f}{f}\quad \Rightarrow \quad W=\frac{\wt{f}\wh{g}}{\wh{f}\wt{g}}\   ,
\end{equation}
in which $g\equiv T_0f=\det(\bun-\bM)$ is the shift in a direction with lattice parameter $a=0$ of the $\tau$-function. Using eqs. 
\eqref{eq:taueq}, which also hold for $g$, together with relations between $f$, $g$ (which follow from \eqref{eq:taueq} taking one or the 
other of the lattice parameters equal to zero)
\bse\label{eq:sgeq}\begin{eqnarray}
&& \ut{g}\wt{f}+\ut{f}\wt{g}=2fg \  , \label{eq:sga} \\ 
&& {\hypohat 0 g} \wh{f}+{\hypohat 0 f} \wh{g}=2fg \  . \label{eq:sgb}
\end{eqnarray}\ese 

Finally, to obtain some explicit formulae for the $\tau$-function \eqref{eq:taudet} we can use the properties of the Cauchy 
matrix $\bM$, in particular the fact that we can explicitly obtain its determinant in factorised form. In fact, invoking 
the  explicit determinantal formula for Cauchy determinants: 
\be\label{eq:cauchydet}
\det(\bA)=\frac{\prod_{i<j}(k_i-k_j)(l_j-l_i)}{\prod_{i,j}(k_i-l_j)}
\ee 
in which $\bA$ is a Cauchy matrix with entries of the form:
$$ A_{i,j}=\frac{1}{k_i-l_j}\quad ,\quad i,j=1,\dots,N\  ,  $$ 
where the $k_i$, $l_j$, $i,j=1,\dots,N$, are a collection of distinct parameters. 
Noting that the matrix $\bM$ can be written as a Cauchy matrix of the form $\bA$ with 
$l_j=-k_j$ multiplied from the left by a diagonal matrix with entries $\rho_i$ and from the right by a diagonal 
matrix with entries $c_j$, we have: 
\begin{equation}\label{eq:Cauchy}
\det\left(\frac{\rho_ic_j}{k_i+k_j}\right)=\left(\prod_i\frac{\rho_ic_i}{2k_i}\right)
\prod_{i<j}\left(\frac{k_i-k_j}{k_i+k_j}\right)^2\  ,  
\end{equation} 
in which $i,j$ can run over any subset of indices of $\{1,\dots,N\}$. 

Furthermore, we have the following expansion formula for the determinant of a matrix of the form ~$\bun+\bM$~
\begin{eqnarray}\label{eq:expand} 
\det\left(\boldsymbol{1}+\boldsymbol{M}\right)&=& 1+\sum_{i=1}^N \left|M_{i,i}\right| 
+\sum_{i<j}\left|\begin{array}{cc} M_{i,i} & M_{i,j} \\ M_{j,i} & M_{j,j} \end{array}\right| \nn \\ 
&& +\sum_{i<j<k}\left|\begin{array}{ccc} M_{i,i} & M_{i,j} & M_{i,k}\\ 
M_{j,i} & M_{j,j} & M_{j,k}\\ M_{k,i} & M_{k,j} & M_{k,k} \end{array}\right| + \cdots + \det(\boldsymbol{M})\  . \nn \\ 
\end{eqnarray}
where as a consequence of \eqref{eq:Cauchy} we can compute all the terms in the expansion \eqref{eq:expand} 
explicitly, leading to a form similar to the celebrated $N$-soliton form for the KdV 
equation given by Hirota, \cite{Hir}.

\section{From lattice KdV soliton solutions to Q3 solitons}
\setcounter{equation}{0}

The main lattice equation arising from the scheme set up in section 2, i.e. equations that we coin ``of KdV type''  is 
the equation \eqref{eq:Uabeq}, which to our knowledge was first given in \cite{NQC}, cf. also \cite{QNCL}, and which we can 
rewrite in affine linear form as follows: 
\begin{equation}\label{eq:seq} 
\begin{split}
\left[1+(p-a)S-(p+b)\wt{S}\right]\left[1+(p-b)\wh{S}-(p+a)\wh{\wt{S}}\right]\\
\quad =\left[1+(q-a)S-(q+b)\wh{S}\right]\left[1+(q-b)\wt{S}-(q+a)\wh{\wt{S}}\right].
\end{split} 
\end{equation}   
Following \cite{RasinNQC}, we will refer to (\ref{eq:seq}) as NQC equation. 
We begin this section by making precise the connection between the NQC equation and the equation ${\rm Q3}_{\delta=0}$. 
This makes some (mainly notational) ground-work for the subsequent statement and proof of the $N$-soliton solution for equation Q3, 
which is constructed on the basis of solutions of \eqref{eq:seq}. In \cite{RasinNQC} a full classification of 
all parameter subcases of \eqref{eq:seq} was given, In particular, it was already noted in the earlier papers that by limits 
on the parameters $a$, $b$ (as either tend to zero or infinity) eq. \eqref{eq:seq} reduces to the other KdV type lattice 
equations, namely \eqref{eq:dKdV}, \eqref{eq:dMKdV} as well as \eqref{eq:dSKdV}. Furthermore, as already remarked in \cite{ABS}, 
\eqref{eq:seq} corresponds to the $\dd=0$ case of Q3. We will first make this connection more explicit, leading to a new 
parametrisation of Q3, and then show that in fact the $N$-soliton solutions of \eqref{eq:seq} for different values of $a$, $b$ 
together constitute a solution for the full case (i.e. $\dd\neq 0$) of Q3, which is the main statement in Theorem 1.

\subsection{Connection between ${\rm Q3}_{\delta=0}$ and the NQC equation} 

We shall now relate eq. \eqref{eq:seq} to the special case of (\ref{eq:Qeqsc}) with $\dd=0$, which we indicate by ${\rm Q3}_{\delta=0}$. 
We first introduce the following dependent variable 
\begin{equation} \label{eq:us}
u^0_{n,m}=\digamma_{n,m}(a,b)\left(1-(a+b)S_{n,m}(a,b)\right)\quad,\quad  
\ee 
in which
\begin{equation}
\digamma_{n,m}(a,b) = \left(\dfrac{P}{(p-a)(p-b)}\right)^n\left(\dfrac{Q}{(q-a)(q-b)}\right)^m\  ,
\label{eq:vpdef}
\end{equation} 
where 
\begin{equation}\label{eq:parcurves}
P^2=(p^2-a^2)(p^2-b^2),\qquad Q^2=(q^2-a^2)(q^2-b^2)\  . 
\end{equation}  
This brings \eqref{eq:seq} in the form: 
\begin{equation} \label{eq:Q30parm} 
P(u^0\wh{u}^0+\wt{u}^0\wh{\wt{u}}^0)-Q(u^0\wt{u}^0+\wh{u}^0\wh{\wt{u}}^0)=
(p^2-q^2)\left(\wh{u}^0\wt{u}^0+u^0\wh{\wt{u}}^0\right)\   , 
\end{equation}  
where the lattice parameters have now become points $\ssp=(p,P)$ respectively $\ssq=(q,Q)$ on the (Jacobi) elliptic curve: 
\begin{equation}
\ssp,\ssq \in \Gamma :=\{(x,X) | X^2=(x^2-a^2)(x^2-b^2) \}.
\label{eq:gamdef}
\end{equation}
The full equation ${\rm Q3}$ now reads 
\begin{equation} \label{eq:Q3parm} 
P(u\wh{u}+\wt{u}\wh{\wt{u}})-Q(u\wt{u}+\wh{u}\wh{\wt{u}})=(p^2-q^2)\left((\wh{u}\wt{u}+u\wh{\wt{u}})+\frac{\dd^2}{4PQ}\right)
\end{equation}  
and this corresponds to the form of the Q3 equation \eqref{eq:Qeqsc} in the ABS list by the following relations between the original parameters 
$\po$, $\qo$ and the new parameters $\ssp$, $\ssq$: 
\begin{equation} \label{eq:origparm} 
\po^2=\frac{p^2-b^2}{p^2-a^2}\quad,\quad P=\frac{(b^2-a^2)\po}{1-\po^2}\quad,\quad  
\qo^2=\frac{q^2-b^2}{q^2-a^2}\quad,\quad Q=\frac{(b^2-a^2)\qo}{1-\qo^2}\  . 
\end{equation} 
while ~$u=(b^2-a^2)\uo$~, the latter being the dependent variable of the equation (\ref{eq:Qeqsc}).

\subsection{N-Soliton structure for Q3}

The main result of the paper \cite{AHN2} was to give the $N$-soliton solution for equation Q3, 
however without presenting there the full proof. Here we will present a constructive proof based on the machinery 
developed in section 2, which we think reveals some of the structures behind the Q3 equation and the solutions, and 
their connection to other lattice equations.  

\paragraph{Theorem 1:} {\it The $N$-soliton solution of ${\rm Q3}$ \eqref{eq:Q3parm}, which we denote by $u^{(N)}=u^{(N)}_{n,m}$ is given by the formula 
\begin{equation}
\label{eq:Q3sol}
\begin{split}
u^{(N)} =&  A\digamma(a,b)\left[ 1-(a+b)S(a,b)\right]+B\digamma(a,-b)\left[ 1-(a-b)S(a,-b)\right]\\ 
& + C\digamma(-a,b)\left[ 1+(a-b)S(-a,b)\right]+ D\digamma(-a,-b)\left[ 1+(a+b)S(-a,-b)\right]
\end{split}
\end{equation}
Here $S(\pm a,\pm b)=S_{n,m}(\pm a,\pm b)$ are the $N$-soliton solutions of the NQC equation \eqref{eq:seq} with parameters $\pm a$, $\pm b$ 
as given in \eqref{eq:Uabs}. 
The function $\digamma(a,b)=\digamma_{n,m}(a,b)$ is defined in \eqref{eq:vpdef} and $A$, $B$, $C$ and $D$ are constants subject to the single constraint:   
\begin{equation}\label{eq:ABCD} 
AD(a+b)^2-BC(a-b)^2=-\frac{\dd^2}{16ab}.
\end{equation}
}
\paragraph{Remark:} The pseudo-linear structure of this solution, as an almost arbitrary linear combination of four different solution of the NQC equation with 
the parameters $a$, $b$ changing signs, is remarkable, the various choices of signs apparently being connected to various choices of pairs of branch 
points of the elliptic curve \eqref{eq:parcurves} of the lattice parameters. As we shall see in the unravelling of this solution, the $a$, $b$ not only 
play the role of moduli of those curves, but as lattice parameters in their own right (w.r.t. additional ``hidden'' lattice directions). 

\noindent
{\bf Proof:} 
We will now go over the various steps needed to prove Theorem 1, all of which are based on the relations of the $N$-soliton solutions to KdV type equations given in section 2.

\paragraph{}{\it Step \# 1.} We first introduce a new associated dependent variable $U^{(N)}=U^{(N)}_{n,m}$ given by:
\begin{equation}
\label{eq:assQ3sol} 
\begin{split}
U^{(N)}=& (a+b)A\digamma(a,b) V(a)\,V(b)+(a-b)B\digamma(a,-b) V(a)\,V(-b)\\ 
& - (a-b)C\digamma(-a,b) V(-a)\,V(b)- (a+b)D\digamma(-a,-b) V(-a)\,V(-b),
\end{split}
\end{equation}
in which $V(\pm a)=V_{n,m}(\pm a)$, $V(\pm b)=V_{n,m}(\pm b)$ are defined by \eqref{eq:Vdef} and $\digamma(a,b)=\digamma_{n,m}(a,b)$ is defined by (\ref{eq:vpdef}).
We recall that the connection 
between the objects $V(a)$, $V(b)$ and $S(a,b)$ is given by \eqref{eq:Uaba}, and that this relation can be covariantly extended to other lattice directions as for example in (\ref{eq:Uabb}). 
Using these with the equations (\ref{eq:UVW})
we can derive the following set of important Miura type relations. 

\paragraph{Lemma 1:} {\it For $u=u^{(N)}_{n,m}$ defined in \eqref{eq:Q3sol} and the associated variable $U=U^{(N)}_{n,m}$ defined in 
\eqref{eq:assQ3sol} the following hold:
\bse\label{eq:Q3Miura}\begin{eqnarray}
p-q+\wh{w}-\wt{w} &=& \frac{1}{U}\left[ P\wt{u}-Q\wh{u}-(p^2-q^2)u\right]  \label{eq:Q3Miuraa} \\    
                    &=& -\frac{1}{\wh{\wt{U}}}\left[ P\wh{u}-Q\wt{u}-(p^2-q^2)\wh{\wt{u}}\right]  \label{eq:Q3Miurab} \\ 
p+q+w-\wh{\wt{w}} &=& \frac{1}{\wh{U}}\left[ P\wh{\wt{u}}-Qu-(p^2-q^2)\wh{u}\right]  \label{eq:Q3Miurac} \\    
                    &=& -\frac{1}{\wt{U}}\left[ Pu-Q\wh{\wt{u}}-(p^2-q^2)\wt{u}\right]  \label{eq:Q3Miurad}  
\end{eqnarray}\ese 
where $w=w^{(N)}_{n,m}$, defined in \eqref{eq:wsol}, is the $N$-soliton solution of the lattice potential KdV equation \eqref{eq:dKdV}.
The relations \eqref{eq:Q3Miura} hold for arbitrary coefficients $A$, $B$, $C$, $D$ (i.e., without invoking the constraint \eqref{eq:ABCD}).}

\noindent{\bf Proof:} This is by direct computation. In fact, to prove \eqref{eq:Q3Miuraa}, we consider 
\begin{eqnarray*}
\fl &&P\wt{u}-Q\wh{u}-(p^2-q^2)u= \\ 
\fl &=& A\digamma(a,b)\left[ (p+a)(p+b)(1-(a+b)\wt{S}(a,b))+\right. \\ 
\fl && \qquad\qquad \left. -(q+a)(q+b)(1-(a+b)\wh{S}(a,b)-(p^2-q^2)(1-(a+b)S(a,b))\right]+ \dots \\
\fl &=& A(a+b)\digamma(a,b)\left[ (p-q)-(p+a)(p+b)\wt{S}(a,b)+(q+a)(q+b)\wh{S}(a,b)+(p^2-q^2)S(a,b)\right]+ \dots \\
\fl &=& A(a+b)\digamma(a,b)\left[ (p+a)\left(1-(p+b)\wt{S}(a,b))+(p-a)S(a,b)\right)+\right. \\ 
\fl && \qquad\qquad\qquad \left. -(q+a)\left(1-(q+b)\wh{S}(a,b)+(q-a)S(a,b)\right)\right]+ \dots \\
\fl &=& A(a+b)\digamma(a,b)\left[ (p+a)\wt{V}(a)V(b)-(q+a)\wh{V}(a)V(b)\right] + \dots \\ 
\fl &=&A(a+b)\digamma(a,b)(p-q+\wh{w}-\wt{w})V(a)V(b) + \dots\\ 
\fl &=& (p-q+\wh{w}-\wt{w})U\  , 
\end{eqnarray*}
in which the dots in each line on the right hand sides stand for similar terms with $(a,b)$ replaced by $(a,-b)$, $(-a,b)$, $(-a,-b)$ and 
$A$ replaced by $B$,$C$,$D$ respectively. In the last steps we have made use of \eqref{eq:Uab} and \eqref{eq:UVWa} respectively. The other relations 
\eqref{eq:Q3Miurab}-\eqref{eq:Q3Miurad} are proven in a similar fashion.\done

\paragraph{}{\it Step \# 2.} We now establish an important property of the objects defined in \eqref{eq:Q3sol} and
\eqref{eq:assQ3sol}, which holds for 
arbitrary coefficients $A$, $B$, $C$, $D$ (i.e., again without invoking the constraint \eqref{eq:ABCD}), namely:

\paragraph{Lemma 2:} {\it The following identities hold for the $N$-soliton expressions:~$u=u^{(N)}_{n,m}$, $U=U^{(N)}_{n,m}$
\bse\label{eq:biquadids}\begin{eqnarray}
&& U\wt{U}-P(u^2+\wt{u}^2)+(2p^2-a^2-b^2)u\wt{u} =\frac{4ab}{P}\det(\cA)\  , \label{eq:biquadidsa} \\ 
&& U\wh{U}-Q(u^2+\wh{u}^2)+(2q^2-a^2-b^2)u\wh{u} =\frac{4ab}{Q}\det(\cA)\  , \label{eq:biquadidsb} 
\end{eqnarray}\ese
in which the $2\times 2$ matrix $\cA$ is given by 
\begin{equation}
\cA=\left(\begin{array}{cc} (a+b)A & (a-b)B \\ -(a-b)C & -(a+b)D \end{array}\right)\   . 
\end{equation}}
{\bf Proof:} Again this is proven by direct computation. However, in order to make the emergence of the determinants more transparent we consider the 
following $2\times 2$ matrices:
\bse\label{eq:Laxadhocs}\begin{eqnarray}
\bL&=:&\left(\begin{array}{ccc} P\wt{u}-(p^2-b^2)u &,& (p-b)U \\ (p+b)\wt{U} &,& -Pu+(p^2-b^2)\wt{u}\end{array}\right)\  , \\ 
\bM&=:&\left(\begin{array}{ccc} Q\wh{u}-(q^2-b^2)u &,& (q-b)U \\ (q+b)\wh{U} &,& -Qu+(q^2-b^2)\wh{u}\end{array}\right)\   .  
\end{eqnarray}\ese     
Evaluating the entries in these matrices we obtain 
\begin{eqnarray}
\fl && P\wt{u}-(p^2-b^2)u = \nn \\ 
\fl &=& A\digamma(a,b)\left[ \frac{P^2}{(p-a)(p-b)}(1-(a+b)\wt{S}(a,b))-(p^2-b^2)(1-(a+b)S(a,b))\right]+ \dots \nn \\ 
\fl &=& A\digamma(a,b)(p+b)\left[ (p+a)(1-(a+b)\wt{S}(a,b))-(p-b)(1-(a+b)S(a,b))\right]+ \dots \nn \\ 
\fl &=& A\digamma(a,b)(p+b)(a+b)\left[ 1-(p+a)\wt{S}(a,b))+(p-b)S(a,b)\right]+ \dots \nn \\ 
\fl &=& A(a+b)\digamma(a,b)(p+b)\wt{V}(b)V(a)+\dots = \sqrt{p^2-b^2}\,\brr^T(a)\cA\wt{\brr}(b)\  , 
\end{eqnarray} 
in which again the dots in each line on the right hand sides stand for similar terms with $(a,b)$ replaced by $(a,-b)$, $(-a,b)$, $(-a,-b)$ and 
$A$ replaced by $B$,$C$,$D$ respectively. On the right-hand side of this expression we have introduced the vectors 
$$\brr(a)=\left( \begin{array}{c} \rho^{1/2}(a)\,V(a) \\ \rho^{1/2}(-a)\,V(-a)\end{array} \right)\quad,\quad 
 \brr^T(b)=\left( \rho^{1/2}(b)\,V(b),\rho^{1/2}(-b)\,V(-b)\right)$$
in which the plane-wave factors $\rho(a)$ are given by 
\begin{equation} \rho(a)=\left(\frac{p+a}{p-a}\right)^n\left(\frac{q+a}{q-a}\right)^m. \label{eq:rho1}\end{equation}
A similar computation as above yields 
\begin{equation}
-Pu+(p^2-b^2)\wt{u}=\sqrt{p^2-b^2}\,\wt{\brr}^T(a)\cA\brr(a)\  , 
\end{equation} 
whilst $U$ and $\wt{U}$ can be written as 
\begin{equation}
U=\brr^T(a)\cA\brr(b)\quad,\quad \wt{U}=\wt{\brr}^T(a)\cA\wt{\brr}(b)\  . 
\end{equation} 
Thus, we find that the matrix $\bL$ in \eqref{eq:Laxadhocs} can be written as
$$  \bL=\left(\begin{array}{ccc} \sqrt{p^2-b^2}\,\brr^T(a)\cA\wt{\brr}(b) &,& (p-b)\brr^T(a)\cA\brr(b)\\ 
(p+b)\wt{\brr}^T(a)\cA\wt{\brr}(b) &,& \sqrt{p^2-b^2}\,\wt{\brr}^T(a)\cA\brr(b) \end{array}\right)\  , 
$$
and similarly for the matrix $\bM$. 
Using now the general determinantal identity
$$  \det\left(\sum_{j=1}^r\,\mbx_j\mby_j^T\right)=\det\left( (\mby_i^T\cdot\mbx_j)_{i,j=1,\cdots,r}\right)  $$ 
for any collection of $r$ pairs of $r$-component column vectors $\mbx_i$,$\mby_i$ (the superindex $T$ denoting transposition), 
we obtain the following result:
\begin{eqnarray*}\label{eq:intermed}
\det(\bL)&=&(p^2-b^2)\det\left(\cA\wt{\brr}(b)\,\brr^T(a)+\cA\brr(b)\,\wt{\brr}^T(a)\right) \\ 
&=&(p^2-b^2)\det(\cA) \det\left(\wt{\brr}(b)\,\brr^T(a)+\brr(b)\,\wt{\brr}^T(a)\right)\\ 
&=& \det\left\{\left(\wt{\brr}(b),\brr(b)\right)\,\left(\begin{array}{c}\brr^T(a)\\ \wt{\brr}^T(a)\end{array}\right)\right\} \\ 
&=& -\det\left(\brr(a),\wt{\brr}(a)\right)\,\det\left(\brr(b),\wt{\brr}(b)\right)\  . 
\end{eqnarray*}
It remains to compute the determinant of the matrix $\left(\brr(a),\wt{\brr}(a)\right)$~ whose columns are the 2-component vectors $\brr(a)$ and $\wt{\brr}(a)$. 
this is done as follows;
\begin{eqnarray*}
\fl \det\left(\brr(a),\wt{\brr}(a)\right)&=& \rho^{1/2}(a)\wt{\rho}^{1/2}(-a)V(a)\,\wt{V}(-a)-\wt{\rho}^{1/2}(a)\rho^{1/2}(-a)\wt{V}(a)\,V(-a) \\ 
\fl &=& \sqrt{\frac{p-a}{p+a}}V(a)\,\wt{V}(-a)-\sqrt{\frac{p+a}{p-a}}\wt{V}(a)\,V(-a)\\
\fl &=& \sqrt{\frac{p-a}{p+a}}\left[ 1-(p+a)\wt{S}(a,-a)+(p+a)S(a,-a)\right]+ \\ 
\fl && \qquad -\sqrt{\frac{p+a}{p-a}}\left[ 1-(p-a)\wt{S}(-a,a)+(p-a)S(-a,a)\right]\\
\fl &=& -\frac{2a}{\sqrt{p^2-a^2}}\   , 
\end{eqnarray*}
where we have used \eqref{eq:Uab} for the choices $b=-a$ and the signs of $a$,$b$ reversed, as well as the fact that $S(a,b)=S(b,a)$. Thus, putting everything 
together we obtain the result:
\begin{equation}\label{eq:detL}
\det(\bL)=(p^2-b^2)\det(\cA)\,\frac{4ab}{\sqrt{p^2-a^2}\,\sqrt{p^2-b^2}}\  . 
\end{equation}
On the other hand a direct computation of the determinant gives:
\begin{eqnarray}\label{eq:detLL}
\det(\bL)&=& -[P\wt{u}-(p^2-b^2)u] [Pu-(p^2-b^2)\wt{u}]-(p^2-b^2)U\wt{U} \nn \\ 
&=& -(p^2-b^2)\left[ P(u^2+\wt{u}^2)-(2p^2-a^2-b^2)u\wt{u}-U\wt{U}\right]\  . 
\end{eqnarray}
Comparing the two expressions for $\det(\bL)$ from \eqref{eq:detL} and \eqref{eq:detLL} we obtain the first equation in Lemma 2. 
\done

\paragraph{}{\it Step \# 3.} The last step is by combining the relations \eqref{eq:Q3Miura} and \eqref{eq:biquadids} as well as the 
potential lattice KdV equation \eqref{eq:dKdV} to assert that $u=u^{(N)}_{n,m}$ solves the Q3 equation. In fact, multiplying for instance  
\eqref{eq:Q3Miurab} by \eqref{eq:Q3Miurad} and using \eqref{eq:biquadidsb}, where we identify 
$$ \det(\cA)=\frac{\dd^2}{16ab}  $$ 
according to \eqref{eq:ABCD}, we obtain from the lattice potential KdV: 
\begin{eqnarray*}  
p^2-q^2 &=& (p+q+w-\wh{\wt{w}})(p-q+\wh{w}-\wt{w}) \\ 
&=& \frac{1}{\wt{U}\wh{\wt{U}}}\left[ P\wh{u}-Q\wt{u}-(p^2-q^2)\wh{\wt{u}}\right]\left[ Pu-Q\wh{\wt{u}}-(p^2-q^2)\wt{u}\right] \\ 
\Rightarrow \\ &&(p^2-q^2)\left[ Q(\wt{u}^2+\wh{\wt{u}}^2)-(2q^2-a^2-b^2)\wt{u}\wh{\wt{u}}+\frac{\dd^2}{4Q}\right] \\ 
&=& P^2(u\wh{u})+Q^2(\wt{u}\wh{\wt{u}}) +(p^2-q^2)^2\wt{u}\wh{\wt{u}} -PQ(u\wt{u}+\wh{u}\wh{\wt{u}}) + \\
&&\qquad -(p^2-q^2)P(\wh{u}\wt{u}+u\wh{\wt{u}}) +(p^2-q^2)Q(\wt{u}^2+\wh{\wt{u}}^2)\  \\ 
&=& P^2(u\wh{u}+\wt{u}\wh{\wt{u}})+(Q^2-P^2)(\wt{u}\wh{\wt{u}}) +(p^2-q^2)^2\wt{u}\wh{\wt{u}} -PQ(u\wt{u}+\wh{u}\wh{\wt{u}}) + \\
&&\qquad -(p^2-q^2)P(\wh{u}\wt{u}+u\wh{\wt{u}}) +(p^2-q^2)Q(\wt{u}^2+\wh{\wt{u}}^2)\  \\ 
\Rightarrow \\ 
&&(p^2-q^2)\left[-(2q^2-a^2-b^2)\wt{u}\wh{\wt{u}}+\frac{\dd^2}{4Q}\right]= \\ 
&=& P\left[ P(u\wh{u}+\wt{u}\wh{\wt{u}})-Q(u\wt{u}+\wh{u}\wh{\wt{u}})-(p^2-q^2)(\wh{u}\wt{u}+u\wh{\wt{u}})\right] + \\ 
&& \quad + (p^2-q^2)\left[ (a^2+b^2-p^2-q^2)\wt{u}\wh{\wt{u}}+(p^2-q^2)\wt{u}\wh{\wt{u}} \right]  
\end{eqnarray*} 
where we have used the fact that $\ssp$, $\ssq$ are on the elliptic curve \eqref{eq:parcurves}  and the identification of $\wt{U}\wh{\wt{U}}$ 
via the result \eqref{eq:biquadids} of Lemma 2. From the last step, after some cancellation of terms, we obtain Q3, in the form \eqref{eq:Q3parm}, 
for the function $u=u^{(N)}_{n,m}$, which completes the proof of the theorem.\DONE

Thus, we have established the explicit form of the $N$-soliton solution \eqref{eq:Q3sol} of the full Q3 equation 
in the form \eqref{eq:Q3parm} in a constructive way, using basically all the ingredients of the lattice KdV type 
soliton structures of section 2. The parametrisation using the KdV lattice parameters $p$, $q$, in terms of which 
the underlying plane-wave factors \eqref{eq:rho} take a simple form, now involving some new fixed parameters $a$, 
$b$, in terms of which we have an elliptic curve \eqref{eq:parcurves} appearing in the equation, of which $\pm a$, 
$\pm b$ are the branch points. It turns out that it is natural now to covariantly extend the lattice to include 
lattice directions for which these parameters $a$, $b$ are the lattice parameters, and thus to introduce 
extensions of the plane-wave factors of the form, \cite{AHN2}: 
\begin{equation}\label{eq:rhoab} 
\rho_{n,m,\alpha,\beta}(k_i):=\left(\frac{p+k_i}{p-k_i}\right)^n
\left(\frac{q+k_i}{q-k_i}\right)^m
\left(\frac{a+k_i}{a-k_i}\right)^\alpha
\left(\frac{b+k_i}{b-k_i}\right)^\beta\rho_i^{0}\ . 
\end{equation}
This allows us then to use relations of the type \eqref{eq:Vtau} and \eqref{eq:Utau} to express the objects 
$V(a)$ and $S(a,b)$ in terms of a $\tau$-function $f=f_{n,m,\alpha,\beta}$ as follows  
\begin{equation}\label{eq:frels}
V(a)=\frac{T_{-a}f}{f}=\frac{T_a^{-1}f}{f}\quad,\quad  
1-(a+b)S(a,b)=\frac{T_{-a}T_{-b}f}{f}=\frac{T_a^{-1}T_b^{-1}f}{f}\    , 
\end{equation}
in which $T_a$, $T_b$ denote elementary shift in the four-dimensional lattice in directions associated with 
the parameters $a$, $b$ (i.e. shifts by one unit in the corresponding lattice variables $\alpha$, $\beta$ 
respectively). Thus, as a consequence of Theorem 1, we get the following 

\paragraph{Corollary:} {\it The solutions \eqref{eq:Q3sol} takes on the following form in terms of the $\tau$-function 
$f=f_{n,m,\alpha,\beta}$, when the plane-wave factors $\rho$ are covariantly extended according to the 
formula \eqref{eq:rhoab}, namely  
\begin{eqnarray} \label{eq:HirQ3sol}
\fl u^{(N)}&=&
 A\digamma(a,b)\,\frac{f_{n,m,\alpha-1,\beta-1}}{f_{n,m,\alpha,\beta}} 
 +B\digamma(a,-b)\, \frac{f_{n,m,\alpha-1,\beta+1}}{f_{n,m,\alpha,\beta}} \nn \\ 
\fl && +C\digamma(-a,b)\, \frac{f_{n,m,\alpha+1,\beta-1}}{f_{n,m,\alpha,\beta}}
+D\digamma(-a,-b)\,\frac{f_{n,m,\alpha+1,\beta+1}}{f_{n,m,\alpha,\beta}} 
\end{eqnarray}
cf. \cite{AHN2}, whilst its adjoint function takes the following form:  
\begin{eqnarray} \label{eq:HirAssQ3sol}
\fl &&U^{(N)}= \nn \\ 
\fl &&=   A(a+b)\digamma(a,b)\,\frac{f_{n,m,\alpha-1,\beta}f_{n,m,\alpha,\beta-1}}{f^2_{n,m,\alpha,\beta}}
+B(a-b)\digamma(a,-b)\, \frac{f_{n,m,\alpha-1,\beta}f_{n,m,\alpha,\beta+1}}{f^2_{n,m,\alpha,\beta}}\nn \\ 
\fl && -C(a-b)\digamma(-a,b)\, \frac{f_{n,m,\alpha+1,\beta}f_{n,m,\alpha,\beta-1}}{f^2_{n,m,\alpha,\beta}}
-D(a+b)\digamma(-a,-b)\,\frac{f_{n,m,\alpha+1,\beta}f_{n,m,\alpha,\beta+1}}{f^2_{n,m,\alpha,\beta}}\  .  \nn \\ 
\fl &&
\end{eqnarray}
} 

As was indicated in \cite{AHN2}, the verification of the solution of Q3 in terms of $f$ can also be done 
by establishing a set of interlinked Hirota-Miwa equations in the four-dimensional space of the variables 
$n,m,\alpha,\beta$. An alternative approach to the $N$-soliton solutions for ABS type equations is developed in 
a subsequent paper, \cite{HieZhang}, in which a formalism is developed in terms of Casorati determinant expressions for 
the $\tau$-functions.

\subsection{Associated biquadratic polynomials and Lax pair} 

In this subsection we make some additional observations and connections related to the $N$-soliton solution for Q3 presented above in Theorem 1.
To this end it will be useful to introduce the following 
\begin{equation}\label{eq:Q3quad}
\cQ_{\ssp,\ssq}(u,\wt{u},\wh{u},\wh{\wt{u}}):= P(u\wh{u}+\wt{u}\wh{\wt{u}})-Q(u\wt{u}+\wh{u}\wh{\wt{u}})-(p^2-q^2)\left((\wh{u}\wt{u}+u\wh{\wt{u}})
+\frac{\dd^2}{4PQ}\,\right)
\end{equation} 
this quadrilateral expression is a polynomial of degree one in four variables with coefficients which depend upon the lattice parameters $\ssp,\ssq\in\Gamma$ (\ref{eq:gamdef}).
By introducing this polynomial the equation (\ref{eq:Q3parm}) may be written conveniently as $\cQ_{\ssp,\ssq}(u,\wt{u},\wh{u},\wh{\wt{u}})=0$.
Following the approach of ABS \cite{ABS} we may associate to this polynomial a biquadratic expression which we denote $\cH$
\begin{equation}
\cH_\ssp(u,\wt{u}):= P(u^2+\wt{u}^2)-(2p^2-a^2-b^2)u\wt{u}+\frac{\dd^2}{4P}.
\label{eq:bqdef}
\end{equation}
This biquadratic is related to (\ref{eq:Q3quad}) in two ways, first, by the equations
\begin{eqnarray} 
&& (p^2-q^2)\cH_\ssp(u,\wt{u})= \cQ_{\ssp,\ssq}(\cQ_{\ssp,\ssq})_{\wh{u}\wh{\wt{u}}}-
(\cQ_{\ssp,\ssq})_{\wh{u}}(\cQ_{\ssp,\ssq})_{\wh{\wt{u}}}   \nn \\ 
&& (q^2-p^2)\cH_\ssq(u,\wh{u})= \cQ_{\ssp,\ssq}(\cQ_{\ssp,\ssq})_{\wt{u}\wh{\wt{u}}}-
(\cQ_{\ssp,\ssq})_{\wt{u}}(\cQ_{\ssp,\ssq})_{\wh{\wt{u}}}   \nn \\ 
\end{eqnarray}
where the subscripts $\wt{u}$ etc. denote partial derivatives.
And, second, by the equation
\begin{equation}\label{eq:biquadeqs}
\cH_\ssp(u,\wt{u})\cH_\ssp(\wh{u},\wh{\wt{u}})-\cH_\ssq(u,\wh{u})\cH_\ssq(\wt{u},\wh{\wt{u}}) 
=\cQ_{\ssp,\ssq}(u,\wt{u},\wh{u},\wh{\wt{u}})\cQ^\ast_{\ssp,\ssq}(u,\wt{u},\wh{u},\wh{\wt{u}})\,
\end{equation}
where $\cQ^\ast_{\ssp,\ssq}$ denotes an associated quadrilateral given by:
\begin{equation}\label{eq:Q3assbiquad}
\cQ^\ast_{\ssp,\ssq}(u,\wt{u},\wh{u},\wh{\wt{u}}):= P(u\wh{u}+\wt{u}\wh{\wt{u}})+Q(u\wt{u}+\wh{u}\wh{\wt{u}})-
(p^2+q^2-a^2-b^2)\left((\wh{u}\wt{u}+u\wh{\wt{u}})-\frac{\dd^2}{4PQ}\,\right)\  . 
\end{equation}
Furthermore, the underlying curve associated to (\ref{eq:Q3quad}), is determined by the polynomial 
in a single variable,
\begin{equation}
r(u)=(\cH_\ssp)_{\wt{u}}^2-2\cH_\ssp(\cH_\ssp)_{\wt{u}\wt{u}}= (a^2-b^2)^2u^2-\dd^2\  . 
\end{equation} 

\vspace{12pt}

The proof in subsection 3.2 relies on relations for a number of quantities that have appeared in passing, 
but which we will now explain somewhat further. In fact, the identities \eqref{eq:biquadids} constitute  
factorisation properties for the aforementioned biquadratics, and can be written as\footnote{Incidentally, 
by performing a natural continuum limit on one of the lattice variables, one can identify the object $U$ 
with a derivative of $u$ with respect to the emerging continuum variable.}:
\be\label{eq:biquadfacts}
\cH_\ssp(u,\wt{u})=U\wt{U}\quad,\quad \cH_\ssq(u,\wh{u})=U\wh{U}\  . 
\ee  
As a consequence of the factorisation \eqref{eq:biquadfacts} and the identity (\ref{eq:biquadeqs}) it becomes clear that either the lattice equation 
~$\cQ_{\ssp,\ssq}(u,\wt{u},\wh{u},\wh{\wt{u}})=0$~ is satisfied, or the equation 
~$\cQ^\ast_{\ssp,\ssq}(u,\wt{u},\wh{u},\wh{\wt{u}})=0$~ is.

\vspace{12pt}

The matrices $\bL$ and $\bM$ appearing in the proof of Lemma 2, cf. \eqref{eq:Laxadhocs}, and whose 
determinants coincide with the biquadratics given above, are 
related to the Lax representation for Q3, which can be written as: 
\bse\label{eq:Lax}\bea
\fl \wt{\bphi}=\bL_\ssk(u,\wt{u})\bphi=\gm_\ssp\left( \begin{array}{ccc}
P\wt{u}-(p^2-k^2)u &,& -K  \\ 
Ku\wt{u}+(p^2-k^2)\frac{\dd^2}{4PK} &,& -Pu+(p^2-k^2)\wt{u}\end{array} 
\right)\bphi\   ,  \\
\fl \wh{\bphi}=\bM_\ssk(u,\wh{u})\bphi=\gm_\ssq\left( \begin{array}{ccc}
Q\wh{u}-(q^2-k^2)u  &,& -K  \\ 
Ku\wh{u}+(q^2-k^2)\frac{\dd^2}{4QK} &,& -Qu+(q^2-k^2)\wh{u}\end{array} 
\right)\bphi\  ,  
\eea\ese 
in which we can set ~$\gamma_\ssp=\cH_\ssp(u,\wt{u})^{-1/2}$~,~$\gamma_\ssq=\cH_\ssq(u,\wh{u})^{-1/2}$~. 
In \cite{BS,Nij} it was explained how to obtain in general Lax representations for quadrilateral lattice equations 
which satisfy the multidimensional consistency property. In the case of the soliton solutions, choosing $u$ as in 
\eqref{eq:Q3sol}, and using the corresponding factorisation of the biquadratics as in \eqref{eq:biquadfacts}, 
with $U$ given by \eqref{eq:assQ3sol}, one can explicitly identify the eigenvector $\bphi$ in terms of the 
ingredients of Theorem 1 and the objects of section 2. In fact, the construction of the Lax representation implies 
that in terms of the components of the vector $\bphi=(\phi_1,\phi_2)^T$ we have that the ratio $\phi_2/\phi_1=T_ku$, 
where it is understood that all the variables are covariantly extended to a lattice involving a lattice direction 
associated with the spectral parameter $k$ as lattice parameter. We will not pursue this identification in the 
present paper.

\section{Connection to the B\"acklund transformation}
\setcounter{equation}{0} 

In this section we study the recursive structure of the $N$-soliton solutions for Q3 presented in Theorem 1, under the application of the  
relevant B\"acklund transformation (BT). Thus, in Theorem 2 we give the precise correspondence between the $N$- and $N+1$-soliton solutions, 
which, as we will see, will involve some redefinitions of the constants as we increase the soliton number. In particular, this is once again 
a manifestation of the multidimensional consistency of the lattice equation, by which we can interpret in a precise sense the lattice equation 
as its own BT. Furthermore, we highlight at the end of this section how the initial step of the B\"acklund chain, i.e. 
the \textit{seed solution} defined as a fixed point of the BT, is embedded in the general formula of Theorem 1.   

\subsection{B\"acklund transform from $N$- to $N+1$-soliton solution} 

The following theorem relates the $N$-soliton solution given previously (\ref{eq:Q3sol}) to the solution one would find by the iterative application of the B\"acklund transformation.

\paragraph{Theorem 2:} {\it Let $u^{(N)}$ be as defined in (\ref{eq:Q3sol}) and let $\wb{u}^{(N+1)}$ be equal to $u^{(N+1)}$ as defined in (\ref{eq:Q3sol}), 
depending on additional parameters $k_{N+1}$, $c_{N+1}$ and additional plane-wave factor $\rho_{N+1}$, and where all but the latter  
plane-wave factors $\rho_i$, $i=1,\dots,N$, as well as the discrete exponentials $\digamma(\pm a,\pm b)$ replaced by  
\be\label{eq:extendrho} 
\ol{\rho}_i=\frac{k_{N+1}+k_i}{k_{N+1}-k_i}\,\rho_i\quad,\quad (i\neq N+1)\quad,\quad 
\wb{\digamma}(\pm a,\pm b)=\frac{K_{N+1}}{(k_{N+1}\mp a)(k_{N+1}\mp b)}\digamma(\pm a,\pm b)\  ,  \ee
respectively. Then $u^{(N)}$ is related to $\ol{u}^{(N+1)}$ by the B\"acklund transformation with 
B\"acklund parameter $\ssk_{N+1}\in\Gamma$,
that is the following equations hold
\begin{equation}\label{eq:Q3solBT} 
\cQ_{\ssp,\ssk_{N+1}}(u^{(N)},\wt{u}^{(N)},\ol{u}^{(N+1)},\wt{\ol{u}}^{(N+1)})=0,\qquad
\cQ_{\ssq,\ssk_{N+1}}(u^{(N)},\wh{u}^{(N)},\ol{u}^{(N+1)},\wh{\ol{u}}^{(N+1)})=0,
\end{equation}
in which $\cQ_{\ssp,\ssk_{N+1}}$, $\cQ_{\ssq,\ssk_{N+1}}$ are the quadrilateral expressions of the form given in 
\eqref{eq:Q3quad}
}
\paragraph{Remark:} The shift $\rho_i\rightarrow\ol{\rho}_i$ corresponds to the covariant extension of the lattice variables in a direction given by the 
new parameter $k_{N+1}$, which equivalently can be described by the introduction of a new discrete lattice variable $h$ associated with 
$k_{N+1}$ as a lattice parameter. This dependence can also be incorporated in the parameters $c_i$, as a hidden dependence on this variable. 
Remarkably, the coefficients $A$, $B$, $C$, $D$, which in principle could alter in the transition from the $N$- to $N+1$-soliton solution, in fact, remain unaltered in the incrementation of the soliton number.  

\noindent 
{\bf Proof:} The proof of Theorem 2 can be broken down into a number of steps, which are all constructive, and rely once again on the machinery developed in section 2. In the first step we break down the $N+1$-soliton expression into components 
associated with the $N$-soliton solution. In the second step we apply the BT to the $N$-soliton solution and in the final 
step we compare the expressions obtained in the two previous steps. 

\paragraph{}{\it Step \# 1.} We first establish a recursive structure between the basic objects like 
$S_{n,m}(a,b)$ and $V_{n,m}(a,b)$ between the $N$- and $N+1$-soliton solutions. This uses the breakdown of 
the Cauchy matrix as it occurs as the kernel $(\bun+\bM)^{-1}$ in the various objects.   

\paragraph{}{Lemma:} The following identity holds for the inverse of a $(N+1)\times(N+1)$ block-matrix 
$$ \left(\begin{array}{c|c} \bA & \bbb \\ \hline  
\bc^T & d\end{array} \right)^{-1} = \left(\begin{array}{c|c} \bA^{-1}(\bun+\frac{1}{s}\bbb\,\bc^T\bA^{-1}) & -\frac{1}{s}\bA^{-1}\bbb \\ 
\hline  -\frac{1}{s}\bc^T\bA^{-1} & \frac{1}{s}\end{array} \right)   $$ 
in which $\bA$ is an invertible $N\times N$ matrix $\bbb$ and $\bc^T$ are $N$-component vector column- and row-vector respectively , 
and $d$ is a nonzero scalar, where the scalar quantity $s$, given by
$$ s=d-\bc^T\bA^{-1}\bbb\  , $$
is assumed to be nonzero.  \\ 
{\bf Proof:} The formula can be verified by direct multiplication. The matrix is invertible if $s$ is nonzero. \done 

Let $\bM^{(N+1)}$ be the $(N+1)\times(N+1)$ Cauchy matrix with parameters $k_1,\dots,k_{N+1}$ as defined in a similar way as 
in \eqref{eq:Cauchymat}. Applying the Lemma to compute the inverse of the matrix ~$\bun+\bM^{(N+1)}$~, which can be decomposed 
as above by setting
$$ \bA=\bun+\bM^{(N)}\quad,\quad \bbb=c_{N+1}(k_{N+1}\bun+\bK)^{-1}\brr\quad,\quad 
\bc=\rho_{N+1}\bc^T(k_{N+1}\bun+\bK)^{-1}\quad,\quad d=1+\frac{\rho_{N+1}c_{N+1}}{2k_{N+1}}\   ,  $$ 
we have 
$$ (\bun+\bM^{(N+1)})^{-1}=\left(\begin{array}{c|c} (\bun+\bM)^{-1}+s^{-1}c_{N+1}\rho_{N+1}\bu(k_{N+1})\tbu(k_{N+1}) & 
-s^{-1}c_{N+1}\bu(k_{N+1}) \\ 
\hline  -s^{-1}\rho_{N+1}\tbu(k_{N+1}) & s^{-1}\end{array} \right) $$ 
where $\bu(\cdot)$ and $\tbu(\cdot)$ are given in \eqref{eq:adefsa} and \eqref{eq:adefsb}, and 
where $s$ now takes the form 
\be\label{eq:s}
 s=1+\frac{\rho_{N+1}c_{N+1}}{2k_{N+1}}\left(1-2k_{N+1}S^{(N)}(k_{N+1},k_{N+1})\right)=\frac{f^{(N+1)}}{f^{(N)}} \   , \ee
where ~$f^{(N+1)}=\det(\bun+\bM^{(N+1)})$~ is the $\tau$-function of the $(N+1)$-soliton solution. 
Using now the definitions \eqref{eq:Uabs} and \eqref{eq:Vdef}, we obtain the following recursion relations for the $N$ to $N+1$ 
soliton objects:
\bse\label{eq:UVsolrecurs}\begin{eqnarray}
\fl S^{(N+1)}(a,b) &=& S^{(N)}(a,b)+\frac{s^{-1}c_{N+1}\rho_{N+1}}{(a+k_{N+1})(b+k_{N+1})}\left(1-(a+k_{N+1})S^{(N)}(a,k_{N+1})\right) 
\times  \nn \\ 
\fl && \qquad\qquad \times \left(1-(b+k_{N+1})S^{(N)}(k_{N+1},b)\right)\   ,  \nn \\
&& \label{eq:UVsola} \\ 
\fl V^{(N+1)}(a) &=& V^{(N)}(a)-\frac{s^{-1}c_{N+1}\rho_{N+1}}{a+k_{N+1}}\left(1-(a+k_{N+1})S^{(N)}(a,k_{N+1})\right) 
V^{(N)}(k_{N+1})\   , \nn \\
&&  \label{eq:UVsolb}  
\end{eqnarray}\ese 
which hold for all $a$, $b$ (not coinciding with $-k_i$, $i=1,\dots,N$). Setting $b=k_{N+1}$ (assumed also not to coincide with any of the 
$-k_i$, $i=1,\dots,N$) we obtain as a corollary of these relations, the following identifications: 
\begin{equation}\label{eq:ss} 
s=\frac{1-(a+k_{N+1})S^{(N)}(a,k_{N+1})}{1-(a+k_{N+1})S^{(N+1)}(a,k_{N})}=\frac{V^{(N)}(k_{N+1})}{V^{(N+1)}(k_{N+1})}\   , 
\end{equation} 
which in particular implies that the ratio of $1-(a+k_{N+1})S(a,k_{N+1})$ between its $N$- and $N+1$-soliton value, is independent 
of the parameter $a$.  

\paragraph{}{\it Step \# 2.}  In this step we will apply the BT with B\"acklund parameter $\ssl=(l,L)\in\Gamma$ to the 
$N$-soliton solution defined in \eqref{eq:Q3sol}, i.e. we want to solve the system of discrete Riccati equations 
for a new variable $v$ 
\be\label{eq:Q3BT}
\cQ_{\ssp,\ssl}(u^{(N)},\wt{u}^{(N)},v,\wt{v})=0\quad, \quad 
 \cQ_{\ssq,\ssl}(u^{(N)},\wh{u}^{(N)},v,\wh{v})=0\  ,  
\ee 
cf. \cite{AHN}. To solve this system we can reduce the problem by identifying two particular solutions, which are 
obtained by covariant extension of the known solution $u^{(N)}$. By covariant extension we mean that all the plane 
wave factors $\rho_i$, ($i=1,\dots,N$), as well as the discrete exponential factors $\digamma_{n,m}(\pm a,\pm b)$,  
which enter in the $N$-soliton formula (e.g. through the Cauchy matrix $\bM=\bM^{(N)}$, and the vectors $\brr$), 
have a dependence on an additional lattice variable $h$ associated with the lattice parameter $l$, 
such that
\bse\label{eq:covext}  
\be\label{eq:rhoextend}
 \rho_i^0=\left(\frac{l+k_i}{l-k_i}\right)^h\rho_i^{00}\quad\Rightarrow\quad 
\ol{\rho}_i=\left(\frac{l+k_i}{l-k_i}\right)\rho_i\   , \ee
and  
\begin{eqnarray}\label{eq:digextend}
&& \digamma_{n,m,h}(\pm a,\pm b)=\digamma_{n,m}(\pm a,\pm b)\left(\frac{L}{(l-a)(l-b)}\right)^h \nn \\  
&&\qquad\Rightarrow\quad \ol{\digamma}(\pm a,\pm b)=\left(\frac{L}{(l\mp a)(l\mp b)}\right)\digamma(\pm a,\pm b)\   .   
\end{eqnarray} \ese 
Here the elementary discrete shift in $h$ is indicated by a $\ol{\phantom{a}}$, which gives meaning to the 
expressions $\ol{u}=T_{l}u$, $\underline{u}=T_l^{-1}u$, i.e. for the thus covariantly extended solution $u^{(N)}_{n,m,h}$ we have 
$$ 
\ol{u}^{(N)}_{n,m,h}=u^{(N)}_{n,m,h+1}\quad,\quad \underline{u}^{(N)}_{n,m,h}=u^{(N)}_{n,m,h-1}\   . 
$$ 
As a consequence of the construction of Theorem 1, the following equations are satisfied:
\bse\label{eq:BT0sol}\begin{eqnarray}\label{eq:BT0sola}
\cQ_{\ssp,\ssl}(u^{(N)},\wt{u}^{(N)},\ol{u}^{(N)},\wt{\ol{u}}^{(N)})=0\quad&,&\quad 
\cQ_{\ssq,\ssl}(u^{(N)},\wh{u}^{(N)},\ol{u}^{(N)},\wh{\ol{u}}^{(N)})=0\   , \\  
\cQ_{\ssp,\ssl}(u^{(N)},\wt{u}^{(N)},\underline{u}^{(N)},\wt{\underline{u}}^{(N)})=0  
\quad&,&\quad  
\cQ_{\ssq,\ssl}(u^{(N)},\wh{u}^{(N)},\underline{u}^{(N)},\wh{\underline{u}}^{(N)})=0\   , \label{eq:BT0solb} 
\end{eqnarray}\ese    
where \eqref{eq:BT0solb} holds because of the symmetry of the quadrilaterals. 
Having established the two solutions $\ol{u}^{(N)}$ and $\underline{u}^{(N)}$ of \eqref{eq:Q3BT} we can now find the 
general solution of that system in the interpolating form:  
\be\label{eq:interpol}
 v=\frac{\ol{u}^{(N)}+\eta\underline{u}^{(N)}}{1+\eta}\   , \ee  
in which $\eta$ is some function to be determined from the following coupled system of homogeneous linear equations:  
\bse\label{eq:etalineqs}\bea
\frac{\wt{\eta}}{\eta} &=& -\frac{\cQ_{\ssp,\ssl}(u^{(N)},\wt{u}^{(N)},\underline{u}^{(N)},\wt{\ol{u}}^{(N)})}
{\cQ_{\ssp,\ssl}(u^{(N)},\wt{u}^{(N)},\ol{u}^{(N)},\wt{\underline{u}}^{(N)})}\   , \\ 
\frac{\wh{\eta}}{\eta} &=& -\frac{\cQ_{\ssq,\ssl}(u^{(N)},\wh{u}^{(N)},\underline{u}^{(N)},\wh{\ol{u}}^{(N)})}
{\cQ_{\ssq,\ssl}(u^{(N)},\wh{u}^{(N)},\ol{u}^{(N)},\wh{\underline{u}}^{(N)})}\   . 
\eea\ese 
The compatibility of this system is equivalent to the compatibility of the BT \eqref{eq:Q3BT} as a coupled system of discrete Riccati equations, which in turn is a consequence of the multidimensional consistency of the lattice 
equation.  
Using the explicit expressions for the quadrilaterals \eqref{eq:Q3quad}, eqs. \eqref{eq:etalineqs} reduce to
\bse\label{eq:etalineqs2}\bea
\frac{\wt{\eta}}{\eta} &=& \frac{P\wt{u}^{(N)}-(p^2-l^2)u^{(N)}-L\underline{u}^{(N)}}{P\wt{u}^{(N)}-(p^2-l^2)u^{(N)}-L\ol{u}^{(N)}}
=\frac{p+l+\underline{w}-\wt{w}}{p-l+\ol{w}-\wt{w}}\   , \\ 
\frac{\wh{\eta}}{\eta} &=& \frac{Q\wh{u}^{(N)}-(q^2-l^2)u^{(N)}-L\underline{u}^{(N)}}{Q\wh{u}^{(N)}-(q^2-l^2)u^{(N)}-L\ol{u}^{(N)}}
=\frac{q+l+\underline{w}-\wh{w}}{q-l+\ol{w}-\wh{w}}\   , 
\eea\ese 
where in the last step we have made use of eqs. \eqref{eq:Q3Miuraa} and the backward shifted version of \eqref{eq:Q3Miurac} in the relevant 
lattice directions. Finally, using eqs. \eqref{eq:UVWb} and \eqref{eq:UVWc} with $q\rightarrow l$ and $\wh{w}\rightarrow\ol{w}$, etc. 
setting $a=l$, which lead to: 
$$ p-l+\ol{w}-\wt{w}=(p-l)\frac{\ol{V}(l)}{\wt{\ol{V}}(l)}\quad, \quad p+l+\underline{w}-\wt{w}=(p+l)\frac{\wt{V}(l)}{V(l)}\  , $$ 
we get 
$$  \frac{\wt{\eta}}{\eta}=\left(\frac{p+l}{p-l}\right)\,\frac{\wt{V}(l)\wt{\ol{V}}(l)}{V(l)\ol{V}(l)}\quad, \quad 
\frac{\wh{\eta}}{\eta}=\left(\frac{q+l}{q-l}\right)\,\frac{\wh{V}(l)\wh{\ol{V}}(l)}{V(l)\ol{V}(l)}\   , $$
which can be simultaneously integrated, yielding the following expression for the function $\eta$:
\be\label{eq:eta} 
\eta=\eta_{n,m}=\eta_0 \left(\frac{p+l}{p-l}\right)^n\left(\frac{q+l}{q-l}\right)^m\,V_{n,m}(l)\ol{V}_{n,m}(l)\  . 
\ee 
Substituting the expression \eqref{eq:eta} into \eqref{eq:interpol} we have obtained the general solution of the 
system \eqref{eq:Q3BT}. It will be convenient to recast the solution \eqref{eq:interpol} in the following form 
\be\label{eq:vsol}
v=\ol{u}^{(N)}-\frac{\eta}{1+\eta}(\ol{u}^{(N)}-\underline{u}^{(N)})=
\ol{u}^{(N)}-\left(\frac{\eta}{1+\eta}\right)\frac{1}{L}(2l+\underline{w}-\ol{w})U^{(N)}\   ,  
\ee 
where we have inserted the expression for the difference ~$\ol{u}^{(N)}-\underline{u}^{(N)}$ which can be 
directly obtained from the backward-shifted relation of \eqref{eq:Q3Miurac} by taking $\ssp=\ssq=\ssl$ and identifying 
the shifts $\wt{w}$ and $\wh{w}$ (and similarly the shifts on $u$ and $U$) with the shift $\ol{w}$ 
associated with the parameter $\ssl$.

\paragraph{Step \# 3.} We now set the B\"acklund parameter $\ssl=\ssk_{N+1}=(k_{N+1},K_{N+1})$, which 
implies that the $\ol{\phantom{a}}$-shift from now on is the lattice shift associated 
with the lattice parameter $\ssk_{N+1}$. This enables us to reexpress the factors in 
\eqref{eq:UVsola} in the following way
\be\label{eq:SSV}
1-(k_{N+1}+a)S^{(N)}(a,k_{N+1})=V^{(N)}(k_{N+1})\underline{V}^{(N)}(a)\   , \ee  
which is an identity that holds for all $a$, and which follows from   
\be\label{eq:lS2}
1+(l-b)\underline{S}^{(N)}(a,b)-(l+a)S^{(N)}(a,b)=\underline{V}^{(N)}(a)V^{(N)}(b)\   , 
\ee 
by setting $l=b=k_{N+1}$, which in turn is a covariantly extended version of \eqref{eq:Uab} 
in the lattice direction associated with parameter $\ssl$.  
Using the identity \eqref{eq:SSV} in \eqref{eq:UVsola} and inserting the result into the expression \eqref{eq:Q3sol} 
for $u^{(N+1)}$ we find  
\begin{eqnarray*}
\fl u^{(N+1)} &=& A\digamma(a,b)\left[1-(a+b)S^{(N+1)}(a,b)\right]+\cdots  \\ 
\fl &=& A\digamma(a,b)\left[ \left(1-(a+b)S^{(N)}(a,b)\right)-
\frac{(a+b)\rho_{N+1}c_{N+1}s^{-1}}{(a+k_{N+1})(b+k_{N+1})}(V^{(N)}(k_{N+1}))^2\underline{V}^{(N)}(a)\underline{V}^{(N)}(b)\right]+ \cdots \\
\fl &=& u^{(N)} - \frac{1}{K_{N+1}}(V^{(N)}(k_{N+1}))^2\rho_{N+1}c_{N+1}s^{-1}
\left[ A(a+b)\underline{\digamma}(a,b)\underline{V}^{(N)}(a)\underline{V}^{(N)}(b)+\cdots \right]\  ,  
\end{eqnarray*}  
where again the dots stand for the remaining terms  with coefficients $B$, $C$, 
$D$ and with $(a,b)$ replaced by $(a,-b)$, $(-a,b)$, $(-a,-b)$ respectively, 
leading to the following formula for the $N+1$-soliton solution:
\be\label{eq:recursSol}
u^{(N+1)}=u^{(N)}-\frac{1}{K_{N+1}}(V^{(N)}(k_{N+1}))^2\rho_{N+1}c_{N+1}s^{-1}\underline{U}^{(N)}\  . 
\ee
On the other hand, setting $\ssl=\ssk_{N+1}$, and choosing the constant $\eta_0=c_{N+1}/(2k_{N+1})$ in \eqref{eq:eta}, 
we obtain the identification 
\be\label{eq:etas} 1+\eta=\ol{s}\   , \ee  
with $s$ given in \eqref{eq:s}, using also \eqref{eq:SSV} in the case of $a=k_{N+1}$. Inserting \eqref{eq:etas} 
into the solution $v$ of the BT, \eqref{eq:vsol}, with $\ssl=\ssk_{N+1}$, and using the relation 
$$ 2k_{N+1}+\ol{w}-\underline{w}= 2k_{N+1}\frac{\ol{V}^{(N)}(k_{N+1})}{V^{(N)}(k_{N+1})}\   ,   $$ 
which follows from the backward shift of \eqref{eq:UVWc} setting $p=q=a=k_{N+1}$,
we can make the identification $v=\ol{u}^{(N+1)}$ between \eqref{eq:vsol} and the forward shift of 
\eqref{eq:recursSol}. This establishes the relation \eqref{eq:Q3solBT}, and hence  
completes the proof of theorem 2. 
\DONE 

Theorem 2 establishes the precise connection between the structure of the $N$-soliton solution as given by the 
Cauchy matrix approach, and the way to generate a soliton hierarchy through B\"acklund transforms. What we conclude 
is that these two approaches coincide up to a subtle identification of the relevant constants in the solution. 
Since, as was remarked in the Corollary of section 3 that the soliton solutions of Q3 really live in an extended 
four-dimensional lattice, the precise identification of those constants is of interest, since they contain possibly 
the additional lattice directions. In fact, in \cite{AHN} we established the first soliton type solutions for Q4 
through the B\"acklund approach, and it is of interest to see how that approach connects to a (yet unknown) 
representation of multi-soliton solutions in terms of a scheme similar to the one set up in this paper for Q3.   

\subsection{Connection to the B\"acklund transformation fixed-point}

In \cite{AHN} we introduced a method for the construction of an elementary solution (of a multidimensionally consistent lattice equation) which is suitable as a seed for the subsequent construction of soliton solutions by iterative application of the B\"acklund transformation. This method is based on an idea which can actually be traced back to Weiss in \cite{wei1}, cf \cite{AN} for a further discussion of this point.
We will now describe the connection between the solution which arises when that method is applied to construct a seed solution for Q3 (\ref{eq:Q3parm}) and the $0$-soliton solution found by substitution of $N=0$ into the principal object studied here (\ref{eq:Q3sol}).
Note that when we substitute $N=0$ in (\ref{eq:Q3sol}) the functions $S(\pm a,\pm b)=S_{n,m}(\pm a,\pm b)$ appearing in the solution all vanish identically.

Following \cite{AHN} we consider a solution of (\ref{eq:Q3parm}) $u_\theta = (u_\theta)_{n,m}$ which is related to itself by the B\"acklund transformation of (\ref{eq:Q3parm}) with B\"acklund parameter $\sst\in\Gamma$ (recall $\Gamma$ was defined in (\ref{eq:gamdef})).
I.e., $u_\theta$ satisfies the system
\begin{equation}
\cQ_{\ssp,\sst}(u_\theta,\wt{u}_\theta,u_\theta,\wt{u}_\theta)=0, \qquad
\cQ_{\ssq,\sst}(u_\theta,\wh{u}_\theta,u_\theta,\wh{u}_\theta)=0.
\label{eq:btfp}
\end{equation}
These are coupled biquadratic equations for $u_\theta$ which are parameter deformations of the biquadratic introduced earlier (\ref{eq:bqdef}).
In this case we need to introduce a limit on the curve, $\Gamma\ni\sst=(t,T)\longrightarrow \infty^+$ in which $t\longrightarrow \infty$ and $T\rightarrow t^2 - a^2/2-b^2/2 +O(t^{-2})$, then the identity
\begin{equation}
\lim_{\sst\longrightarrow \infty^+} \cQ_{\ssp,\sst}(u,\wt{u},u,\wt{u}) = -\cH_\ssp(u,\wt{u}),
\end{equation}
reveals the sense in which the one is a deformation of the other. 
Recall \cite{AHN} that with this special choice of $\sst$ the system (\ref{eq:btfp}) defines a {\it non-germinating} seed solution (or {\it singular} solution in \cite{ABS2}).

To solve simultaneously the biquadratic equations (\ref{eq:btfp}) (for general $\sst$) we introduce parameters $p_\theta$ and $q_\theta$ defined in terms of $\ssp$ and $\ssq$ by the quadratic equations
\begin{equation}
p_\theta + 1/p_\theta = 2\frac{T-t^2+p^2}{P}, \qquad q_\theta + 1/q_\theta = 2\frac{T-t^2+q^2}{Q}.
\end{equation}
The (canonical) solution of (\ref{eq:btfp}) may then be written
\begin{equation}
u_\theta = (u_\theta)_{n,m} = A_\theta p_\theta^n q_\theta^m +B_\theta p_\theta^{-n} q_\theta^{-m},
\label{eq:seed}
\end{equation}
where the coefficients $A_\theta$ and $B_\theta$ are constants subject to the single constraint
\begin{equation}
A_\theta B_\theta = \frac{\delta^2}{16T(2t^2-2T-a^2-b^2)}.
\label{eq:AB}
\end{equation}

We are now in a position to describe the connection between the solution found as a fixed-point of the B\"acklund transformation (\ref{eq:seed}) and the solution $u^{(0)}$, i.e., the solution obtained by setting $N=0$ in (\ref{eq:Q3sol}).
Whereas that solution contains four terms with coefficients $A$, $B$, $C$ and $D$, the solution (\ref{eq:seed}) contains only two terms, so it is clear the solutions do not coincide.
Actually what we find is that making the particular choice $\sst=(0,ab)$ the solution (\ref{eq:seed}) becomes $u^{(0)}$ with $A=A_\theta$, $D=B_\theta$ and $B=C=0$, whilst making the choice $\sst=(0,-ab)$ (\ref{eq:seed}) becomes $u^{(0)}$ with $A=D=0$, $B=A_\theta$ and $C=B_\theta$. In both cases the constraint (\ref{eq:AB}) becomes (\ref{eq:ABCD}).

\section{$N$-soliton solutions for the degenerate sub cases of Q3}
\setcounter{equation}{0} 

The coalescence scheme illustrated in figure \ref{fig:QH} will be used in this section to construct $N$-soliton solutions for the equations Q2, Q1, H3, H2 and H1 by degeneration from the $N$-soliton solution we have given for Q3.
We begin by detailing how the equations themselves are found by degeneration, then we show how to find the new solutions.
\begin{figure}[h]
\xymatrix{
&&\framebox{Q3} \ar[dd] \ar[rr] && \framebox{Q2} \ar[dd] \ar[rr] && \framebox{Q1} \ar[dd] \\ \\
&&\framebox{H3} \ar[rr] && \framebox{H2} \ar[rr] && \framebox{H1}
}
\caption{
Coalescence scheme employed to construct $N$-soliton solutions for the degenerate sub cases of equation Q3.
\label{fig:QH}
}
\end{figure}
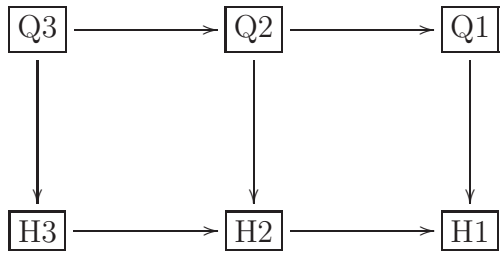

\subsection{The degenerations from Q3}
We degenerate from Q3 as parametrised in (\ref{eq:Q3parm}) which differs from the parametrisation given originally by ABS \cite{ABS} (which we reproduce in (\ref{eq:Qeqsc})).
The degenerate equations consequently appear in a parametrisation different than in the lists (\ref{eq:Qeqs}) and (\ref{eq:Heqs}).
Importantly, throughout the scheme depicted in figure \ref{fig:QH} there is no limit taken on the parameters $p$ and $q$ present in (\ref{eq:Q3parm}), so each equation emerges in terms of these {\it common} lattice parameters (which later will be seen to occur naturally in all of the $N$-soliton solutions).
These parameters are simply related to the parameters $\po$ and $\qo$ of the equations listed in (\ref{eq:Qeqs}) and (\ref{eq:Heqs}) by the following associations
\bse
\begin{eqnarray}
Q3: \ & \po = \dfrac{P}{p^2-a^2} = \dfrac{p^2-b^2}{P}, \quad & \qo = \dfrac{Q}{q^2-a^2} = \dfrac{q^2-b^2}{Q},\label{eq:ppQ3}\\
Q2,Q1: \ & \po = \dfrac{a^2}{p^2-a^2}, & \qo=\dfrac{a^2}{q^2-a^2},\label{eq:ppQ12}\\
H3: \ & \po = \dfrac{P}{a^2-p^2} = \dfrac{1}{P}, & \qo = \dfrac{Q}{a^2-q^2} = \dfrac{1}{Q},\label{eq:ppH3}\\
H2,H1: \ & \po = -p^2, & \qo = -q^2.\label{eq:ppH12}
\end{eqnarray}
\label{eq:paramlist}\ese
When written in terms of these common parameters $p$ and $q$, the equations Q3 and H3 involve also upper-case parameters $P$ and $Q$ which are defined in terms of $p$ and $q$ by an algebraic relation.
For Q3 this relation was introduced previously (\ref{eq:parcurves}) and is given again in (\ref{eq:ppQ3}), for H3 there is a different algebraic relation which is given in (\ref{eq:ppH3}).

It is by limits on the parameters $a$ and $b$ and on the dependent variable $u$ that the degenerations appearing in figure \ref{fig:QH} are achieved.
Specifically, the following list of substitutions result in the indicated degenerations in the limit $\epsilon\longrightarrow 0$:
\bse\begin{eqnarray}
Q3 \longrightarrow Q2: \ & b=a(1-2\epsilon), &\ u\rightarrow \dfrac{\delta}{4a^2}\left(\dfrac{1}{\epsilon}+1+(1+2u)\epsilon\right),\label{eq:Q3Q2}\\
Q2 \longrightarrow Q1: \ &&\ u \rightarrow \dfrac{\delta^2}{4\epsilon^2}+\dfrac{1}{\epsilon}u,\label{eq:Q2Q1}\\
Q3 \longrightarrow H3: \ & b=\dfrac{1}{\epsilon^2}, &\ u\rightarrow \dfrac{\sqrt{\delta}}{2}\epsilon^3 u,\label{eq:Q3H3}\\
Q2 \longrightarrow H2: \ & a=\dfrac{1}{\epsilon}, &\ u\rightarrow \dfrac{1}{4}+\epsilon^2 u,\label{eq:Q2H2}\\
Q1 \longrightarrow H1: \ & a=\dfrac{1}{\epsilon}, &\ u\rightarrow \epsilon \delta u,\label{eq:Q1H1}\\
H3 \longrightarrow H2: \ & a=\dfrac{1}{\epsilon^2}, &\ u \rightarrow \sqrt{-\delta}\epsilon\left(1+\dfrac{\epsilon^4}{2}u\right),\label{eq:H3H2}\\
H2 \longrightarrow H1: &&\ u \rightarrow \dfrac{1}{\epsilon^2} + \dfrac{2}{\epsilon}u.\label{eq:H2H1}
\end{eqnarray}
\label{eq:degenlist}\ese
Note that for these degenerations we assume the parameter $\delta$ appearing in the equations Q3, Q1 and H3 is nonzero.

\subsection{$N$-soliton solutions of Q2 and Q1}
We degenerate from the previously established $N$-soliton solution for Q3 (\ref{eq:Q3sol}). 
To begin we will consider in detail the degeneration from this solution to the $N$-soliton solution of Q2. 
We are led by the requirement that the parameter $b$ which appears in this solution should, according to (\ref{eq:Q3Q2}), be replaced with $b=a(1-2\epsilon)$.
Making this substitution and expanding the result in powers of $\epsilon$ results in a lengthy expansion of (\ref{eq:Q3sol}) which we break down into expansions of the simpler component parts.
First we consider the function $\digamma(a,b)$ defined in (\ref{eq:vpdef}) which has the following expansions depending on its choice of argument
\begin{equation}
\begin{array}{rl}
\digamma(a,b)   \longrightarrow& \rho( a)(1+\epsilon\xi+\epsilon^2(\xi^2/2+\chi)) + O(\epsilon^3),\\
\digamma(a,-b) \longrightarrow& 1-\epsilon\xi+\epsilon^2(\xi^2/2-\chi) + O(\epsilon^3),\\
\digamma(-a,b) \longrightarrow& 1+\epsilon\xi+\epsilon^2(\xi^2/2+\chi) + O(\epsilon^3),\\
\digamma(-a,-b) \longrightarrow& \rho(-a)(1-\epsilon\xi+\epsilon^2(\xi^2/2-\chi)) + O(\epsilon^3),\\
\end{array}
\label{eq:vpexp}
\end{equation}
in which we have introduced the new functions
\begin{equation}
\begin{array}{l}
\xi=\xi_{n,m}=2a\left(\dfrac{p}{a^2-p^2}n+\dfrac{q}{a^2-q^2}m\right), \\ 
\chi=\chi_{n,m}=4a^3\left(\dfrac{p}{(a^2-p^2)^2}n+\dfrac{q}{(a^2-q^2)^2}m\right),
\label{eq:xidef}
\end{array}
\end{equation}
and the function $\rho(a)=\rho_{n,m}(a)$ coincides with the one defined previously (\ref{eq:rho1}).
In a similar fashion we find the expansion of terms involving $\UM(a,b)$ defined in (\ref{eq:adefs}) to be the following
\begin{equation}
\begin{array}{rl}
1-(a+b)\UM(a,b) \longrightarrow& 1-2a\UM(a,a)+O(\epsilon),\\
1-(a-b)\UM(a,-b) \longrightarrow& 1-\epsilon 2a\UM(a,-a)+\epsilon^24a^2\UU(a,-a)+O(\epsilon^3),\\
1+(a-b)\UM(-a,b) \longrightarrow& 1+\epsilon 2a\UM(-a,a)+\epsilon^24a^2\UU(-a,a)+O(\epsilon^3),\\
1+(a+b)\UM(-a,-b) \longrightarrow& 1+2a\UM(-a,-a)+O(\epsilon),\\
\end{array}
\label{eq:Uexp}
\end{equation}
in which we have introduced the new quantity
\begin{equation}
\UU(a,b)=\tbc\,(b\bun+\bK)^{-2}\bu(a)=\tbc\,(b\bun+\bK)^{-2}\,(1+\bM)^{-1}\,(a\bun+\bK)^{-1}\,\brr.\label{eq:UUdef}
\end{equation}
Now, although the expansions of these components of (\ref{eq:Q3sol}) are fixed, by choosing the dependence of the constants $A$, $B$, $C$ and $D$ in (\ref{eq:Q3sol}) on the small parameter $\epsilon$ we can exert some control over its overall expansion.
In order that it be of the required form, namely $u^{(N)}\longrightarrow \dfrac{\delta}{4a^2}\left(\dfrac{1}{\epsilon}+1+(1+2u^{(N)})\epsilon\right)$ as listed in (\ref{eq:Q3Q2}), we make the following choices for these constants
\begin{equation}
\begin{array}{l}
A \rightarrow \dfrac{\delta}{4a^2}A\epsilon,\\
B \rightarrow \dfrac{\delta}{8a^2}\left(\dfrac{1}{\epsilon}+1-\xi_0+((3+\xi_0^2)/2+2AD)\epsilon\right),\\
C \rightarrow \dfrac{\delta}{8a^2}\left(\dfrac{1}{\epsilon}+1+\xi_0+((3+\xi_0^2)/2+2AD)\epsilon\right),\\
D \rightarrow \dfrac{\delta}{4a^2}D\epsilon,\\
\end{array}
\label{eq:Q2const}
\end{equation}
Here the four constants $A$, $B$, $C$ and $D$ (constrained by (\ref{eq:ABCD})) which are present in the Q3 $N$-soliton solution (\ref{eq:Q3sol}) have been replaced with three modified constants $A$, $D$ and $\xi_0$ (with no constraint), in the degeneration to the Q2 $N$-soliton solution.
The number of these constants minus the number of constraints is therefore preserved in the degeneration, which is strong evidence that the solution found by degeneration is the most general one obtainable by this method.
Finally then, the solution of Q2 which arises in the limit $\epsilon\longrightarrow 0$ as a consequence of the substitutions (\ref{eq:vpexp}), (\ref{eq:Uexp}) and (\ref{eq:Q2const}) into (\ref{eq:Q3sol}) is found to be
\begin{equation}
\begin{split}
u^{(N)} =& \dfrac{1}{4}((\xi+\xi_0)^2+1)+a(\xi+\xi_0)\UM(-a,a)+a^2\left(\UU(a,-a)+\UU(-a,a)\right) + \\ 
& AD + \dfrac{1}{2}A\rho(a)(1-2a\UM(a,a))+\dfrac{1}{2}D\rho(-a)(1+2a\UM(-a,-a)).
\end{split}
\label{eq:Q2sol}
\end{equation}
Here $\xi_0$, $A$ and $D$ are the aforementioned constants which may be chosen arbitrarily, $\xi=\xi_{n,m}$ and $\rho(a)=\rho_{n,m}(a)$ are defined in (\ref{eq:xidef}) and (\ref{eq:rho1}), and $\UM(a,b)=\UM_{n,m}(a,b)$ and $\UU(a,b)=\UU_{n,m}(a,b)$ are defined in (\ref{eq:adefs}) and (\ref{eq:UUdef}) respectively.

We now consider the degeneration $Q2\longrightarrow Q1$ which is somewhat simpler than the $Q3\longrightarrow Q2$ degeneration considered above.
To achieve the required limit (\ref{eq:Q2Q1}) of the Q2 $N$-soliton solution (\ref{eq:Q2sol}), namely that $u^{(N)}\longrightarrow\dfrac{\delta^2}{4\epsilon^2}+\dfrac{1}{\epsilon}u^{(N)}$, we need only substitute the constants appearing in the solution as follows
\begin{equation}
\begin{array}{l}
A \rightarrow \dfrac{2A}{\epsilon},\\
D \rightarrow \dfrac{2D}{\epsilon},\\
\xi_0 \rightarrow \xi_0 + \dfrac{2B}{\epsilon}.
\end{array}
\label{eq:Q1const}
\end{equation}
The desired limit results from substitution of (\ref{eq:Q1const}) into (\ref{eq:Q2sol}), provided the modified constants $A$, $B$, $D$ and $\xi_0$ are chosen to satisfy the single constraint
\begin{equation} AD +\dfrac{1}{4}B^2=\dfrac{\delta^2}{16}.\end{equation}
Notice that again the number of constants minus constraints is preserved in the Q2 $\longrightarrow$ Q1 degeneration.
The $N$-soliton solution of Q1 which emerges is
\begin{equation}
u^{(N)} = A\rho(a)(1-2a\UM(a,a)) + B(\xi+\xi_0+2a\UM(-a,a)) + D\rho(-a)(1+2a\UM(-a,-a)).
\label{eq:Q1sol}
\end{equation}

As a small side development here we now consider the further limit $a\longrightarrow 0$ in the Q1 $N$-soliton solution (\ref{eq:Q1sol}).
There is some subtlety in this limit and the solution of Q1 which emerges inspires some useful additional observations.
Performing the $a\longrightarrow 0$ limit naively would also change the equation (in fact send it to Q1$_{\delta=0}$) because $a$ appears in its parametrisation (\ref{eq:ppQ12}).
However, the substitution
\begin{equation}
a=\epsilon, \qquad u\rightarrow 1+\epsilon^2 u,
\end{equation}
preserves the full equation as $\epsilon \longrightarrow 0$. This leads to a reparametrisation of Q1 in which we identify the parameters $\po$ and $\qo$ present in (\ref{eq:Qeqsa}) simply as
\begin{equation}
\po = \dfrac{1}{p^2}, \quad \qo = \dfrac{1}{q^2},
\end{equation}
which come to replace the associations in (\ref{eq:ppQ12}).
Now the same substitution, $a=\epsilon$, yields the following small-$\epsilon$ expansion for the component parts of the solution (\ref{eq:Q1sol}),
\begin{equation}
\begin{array}{rl}
\rho(a) \longrightarrow & 1+\epsilon \nu + \epsilon^2\frac{\nu^2}{2} + O(\epsilon^3),\\
\rho(-a) \longrightarrow & 1-\epsilon \nu + \epsilon^2\frac{\nu^2}{2} + O(\epsilon^3),\\
\xi \longrightarrow & -\epsilon\nu +O(\epsilon^3),\\
\UM(a,a) \longrightarrow & \UD{-1}{-1} - 2\epsilon \UD{-1}{-2} + O(\epsilon^2),\\
\UM(-a,a) \longrightarrow & \UD{-1}{-1} + O(\epsilon^2),\\
\UM(-a,-a) \longrightarrow & \UD{-1}{-1} + 2\epsilon \UD{-1}{-2} + O(\epsilon^2),
\end{array}
\label{eq:Q1exp2}
\end{equation}
where we have introduced the new function $\nu$,
\begin{equation}
\nu=\nu_{n,m}=\dfrac{2}{p}n + \dfrac{2}{q}m.
\label{eq:nudef}
\end{equation}
If we also make the choice
\begin{equation}
\begin{array}{rl}
A\rightarrow &\dfrac{1}{2} + \dfrac{\epsilon}{2}(A+\nu_0),\\
B\rightarrow &1-\dfrac{\delta}{2} + \dfrac{\epsilon}{2}((1-\delta)\nu_0-2A\nu_1/\nu_0),\\
D\rightarrow &\dfrac{\delta-1}{2} + \dfrac{\epsilon}{2}A(2\nu_1/\nu_0-1),\\
\xi_0 \rightarrow & 1 - \epsilon \nu_0 + \dfrac{\epsilon^2}{2} \nu_0^2,
\end{array}
\end{equation}
for the constants appearing in (\ref{eq:Q1sol}) then the substitution of (\ref{eq:Q1exp2}) yields the desired limit of the solution $u^{(N)}\longrightarrow 1+\epsilon^2 u^{(N)}$.
The resulting solution of Q1 is as follows
\begin{equation}
u^{(N)}=\delta\left(\dfrac{1}{4}(\nu+\nu_0)^2-(\nu+\nu_0)\UD{-1}{-1}+2\UD{-1}{-2}\right) + A(\nu +\nu_1 -2\UD{-1}{-1}),
\label{eq:Q1sol2}
\end{equation}
where the constants $\nu_0$, $\nu_1$ and $A$ are arbitrary, $\nu=\nu_{n,m}$ is defined in (\ref{eq:nudef}) and $S^{(i,j)}=S^{(i,j)}_{n,m}$ is defined in (\ref{eq:U}).

The main point we wish to make about the Q1 solution (\ref{eq:Q1sol2}) is that it generalizes the solution previously given (\ref{eq:zsol}) for the lattice Schwarzian KdV equation, i.e., the equation Q1$_{\delta=0}$.
Specifically this can be seen as an extension of that solution to the case $\delta\neq 0$ in that it reduces to (\ref{eq:zsol}) if we take $\delta=0$ and $A=-1/2$.

\subsection{$N$-soliton solution of H3}
To find the $N$-soliton solution for H3 we degenerate from the Q3 $N$-soliton solution (\ref{eq:Q3sol}) led now by the requirement that, according to (\ref{eq:Q3H3}), we choose $b=\dfrac{1}{\epsilon^2}$.
As before we give small-$\epsilon$ expansions for the component parts of the Q3 $N$-soliton solution (\ref{eq:Q3sol}) which result from making this substitution for $b$.
We find that
\begin{equation}
\begin{array}{rl}
\digamma(a,b)   \longrightarrow& \vartheta+ O(\epsilon^2),\\
\digamma(a,-b) \longrightarrow& (-1)^{n+m}\vartheta+ O(\epsilon^2),\\
\digamma(-a,b) \longrightarrow& (-1)^{n+m}\vartheta^{-1} + O(\epsilon^2),\\
\digamma(-a,-b) \longrightarrow& \vartheta^{-1} + O(\epsilon^2),\\
1-(a+b)\UM(a,b) \longrightarrow& V(a)+O(\epsilon^2),\\
1-(a-b)\UM(a,-b) \longrightarrow& V(a)+O(\epsilon^2),\\
1+(a-b)\UM(-a,b) \longrightarrow& V(-a)+O(\epsilon^2),\\
1+(a+b)\UM(-a,-b) \longrightarrow& V(-a)+O(\epsilon^2),
\end{array}
\label{eq:expH3}
\end{equation}
where we have introduced the new function
\begin{equation}
\vartheta=\vartheta_{n,m} = \left(\frac{P}{a-p}\right)^n\left(\frac{Q}{a-q}\right)^m,
\label{eq:vthdef}
\end{equation}
which involves parameters $P$ and $Q$ which are related to $p$ and $q$ by (\ref{eq:ppH3}), and where $V(a)=V_{n,m}(a)$ is defined in (\ref{eq:Vdef}).

Substituting the expressions (\ref{eq:expH3}) into (\ref{eq:Q3sol}) whilst choosing the constants in that solution to be
\begin{equation}
\begin{array}{ll}
A\rightarrow \epsilon^3\dfrac{\sqrt{\delta}}{2}A,&
B\rightarrow \epsilon^3\dfrac{\sqrt{\delta}}{2}B,\\
C\rightarrow \epsilon^3\dfrac{\sqrt{\delta}}{2}C,&
D\rightarrow \epsilon^3\dfrac{\sqrt{\delta}}{2}D,\\
\end{array}
\label{eq:H3const}
\end{equation}
we find $u^{(N)}\longrightarrow \epsilon^3\dfrac{\sqrt{\delta}}{2}u^{(N)}$ as required for the Q3$\longrightarrow$H3 limit given in (\ref{eq:Q3H3}).
Thus we find the $N$-soliton solution of H3 to be 
\begin{equation}\label{eq:H3sol}
u^{(N)}=(A+(-1)^{n+m}B)\vartheta V(a) + ((-1)^{n+m}C+D)\vartheta^{-1}V(-a).
\end{equation}
Here the constants $A$, $B$, $C$ and $D$ are subject to the single constraint
$$ AD-BC = \dfrac{-\delta}{4a}, $$
which follows by substitution of $b=\dfrac{1}{\epsilon^2}$ and (\ref{eq:H3const}) into (\ref{eq:ABCD}).
The functions $\vartheta=\vartheta_{n,m}$ and $V(a)=V_{n,m}(a)$ are defined in (\ref{eq:vthdef}) and (\ref{eq:Vdef}).

\subsection{$N$-soliton solutions of H2 and H1}
To find the H2 $N$-soliton solution we choose to degenerate from the Q2 $N$-soliton solution (\ref{eq:Q2sol}) led by the requirement that we substitute $a=\dfrac{1}{\epsilon}$ (cf \eqref{eq:Q2H2}). (Observe from figure \ref{fig:QH} that we could choose to degenerate from the H3 $N$-soliton solution (\ref{eq:H3sol}) to the H2 $N$-soliton solution, the two paths actually lead to the same result.)
Making the substitution $a=\dfrac{1}{\epsilon}$ into the component parts of (\ref{eq:Q2sol}) yields the following small-$\epsilon$ expansions:
\begin{equation}
\begin{array}{rl}
\xi \longrightarrow& \epsilon \zeta + O(\epsilon^3),\\
\rho(a) \longrightarrow&(-1)^{n+m}(1+\epsilon \zeta + \epsilon^2\zeta^2/2 + O(\epsilon^3)),\\ 
\rho(-a) \longrightarrow&(-1)^{n+m}(1-\epsilon \zeta + \epsilon^2\zeta^2/2 + O(\epsilon^3)),\\
\end{array}
\label{eq:H2sub1}
\end{equation}
in which we have introduced a new function $\zeta$,
\begin{equation}
\zeta=\zeta_{n,m}=2np+2mq,
\label{eq:zdef}
\end{equation}
and
\begin{equation}
\begin{array}{rl}
a\UM(-a,a) \longrightarrow & -\epsilon \UD{0}{0} + O(\epsilon^2),\\
a\UM(a,a) \longrightarrow & \epsilon \UD{0}{0}-2\epsilon^2 \UD{0}{1} + O(\epsilon^3),\\
a\UM(-a,-a) \longrightarrow & \epsilon \UD{0}{0}+2\epsilon^2 \UD{0}{1} + O(\epsilon^3),\\
a^2(\UU(-a,a)+\UU(a,-a)) \longrightarrow & 2\epsilon^2 \UD{0}{1} + O(\epsilon^3).
\label{eq:H2sub2}
\end{array}
\end{equation}
Substituting (\ref{eq:H2sub1}) and (\ref{eq:H2sub2}) into (\ref{eq:Q2sol}) combined with the following choice for the constants
\begin{equation}
\begin{array}{rl}
A\rightarrow & A(\epsilon+\zeta_1\epsilon^2/2),\\
D\rightarrow & A(-\epsilon+\zeta_1\epsilon^2/2),\\
\xi_0\rightarrow & \epsilon \zeta_0,
\end{array}
\end{equation}
results in an expansion for (\ref{eq:Q2sol}) of the required form $u^{(N)}\longrightarrow\frac{1}{4}+\epsilon^2u^{(N)}$ (cf (\ref{eq:Q2H2})) with the new H2 $N$-soliton solution which results being
\begin{equation}\label{eq:H2sol}
u^{(N)} = \dfrac{1}{4}(\zeta+\zeta_0)^2 -(\zeta+\zeta_0) \UD{0}{0} + 2\UD{0}{1}-A^2+(-1)^{n+m}A(\zeta+\zeta_1-2\UD{0}{0}),
\end{equation}
where the constants $A$, $\zeta_0$ and $\zeta_1$ are arbitrary (and unrelated),
$\zeta=\zeta_{n,m}$ is defined in (\ref{eq:zdef}) and $S^{(i,j)}=S^{(i,j)}_{n,m}$ is defined in (\ref{eq:U}).

We remark that the solution (\ref{eq:H2sol}) in the case $A=0$ can be transformed to the solution given previously for the equation Q1 (\ref{eq:Q1sol2}) with $A=0$ and $\delta=1$ by the simple (self inverse) transformation
\begin{equation}
p,q,k_1\ldots k_N \rightarrow 1/p,1/q,1/k_1\ldots 1/k_N, \qquad \rho_i\rightarrow (-1)^{n+m}\rho_i.
\end{equation}
The connection between these solutions reflects a kind of duality between the equations Q1$_{\delta=1}$ and H2 which was found previously in \cite{Atkinson}.

To find the $N$-soliton solution for the equation H1 we choose to degenerate from the $N$-soliton solution of Q1 (\ref{eq:Q1sol}).
According to (\ref{eq:Q1H1}) we should make the substitution $a=\dfrac{1}{\epsilon}$ into (\ref{eq:Q1sol}), conveniently we have already expanded the component parts of this solution in powers of $\epsilon$ because they appeared in our consideration of the Q2 $\longrightarrow$ H2 degeneration detailed above in (\ref{eq:H2sub1}) and (\ref{eq:H2sub2}).
The required limit of the solution, which according to (\ref{eq:Q1H1}) is $u^{(N)} \longrightarrow \epsilon\delta u^{(N)}$, is achieved by choosing the constants appearing in (\ref{eq:Q1sol}) as follows:
\begin{equation}
\begin{array}{rl}
A\rightarrow & \dfrac{\delta}{2} A(1+\zeta_1\epsilon),\\
D\rightarrow & \dfrac{\delta}{2} A(-1+\zeta_1\epsilon),\\
B\rightarrow & \delta B,\\
\xi_0 \rightarrow & \epsilon \zeta_0.
\end{array}
\end{equation}
The resulting $N$-soliton solution of H1 reads
\begin{equation}\label{eq:H1sol}
u^{(N)} = B(\zeta+\zeta_0-2\UD{0}{0})+(-1)^{n+m}A(\zeta+\zeta_1-2\UD{0}{0}),
\end{equation}
where $\zeta_0$, $\zeta_1$, $A$ and $B$, subject to the single constraint
$$ A^2-B^2=\dfrac{-1}{4}, $$
are otherwise arbitrary constants, $\zeta=\zeta_{n,m}$ and $\UD{i}{j}=\UD{i}{j}_{n,m}$ are defined in (\ref{eq:zdef}) and (\ref{eq:U}) respectively.

\subsection{Equations A2 and A1}
In the above we have constructed solutions by degeneration following the coalescence diagram of figure \ref{fig:QH}. 
This diagram does not include the equations A2 and A1 and we have thus far not given explicitly their $N$-soliton solutions.
However these equations are related to Q3$_{\delta=0}$ and Q1 respectively by straightforward gauge transformation, so the solutions we have given for those equations may be transformed to solutions for the equations A2 and A1.

\section{Concluding remarks}
\setcounter{equation}{0} 

In this paper we have reviewed the construction of $N$-soliton solution for integrable quadrilateral lattice equations 
of KdV type dating back to the early 1980s, cf. \cite{NQC,QNCL}, as well as constructed  
$N$-soliton solutions for the majority of equations in the ABS list. We have concentrated particular on the case of 
${\rm Q3}$, which by degeneration yields all the other ABS equations (except Q4) by limits on the parameters. The 
$N$-soliton solution of Q3 is particularly interesting as it is most conveniently described in a four-dimensional 
lattice, associated with four lattice parameters, two of which are the lattice parameters of the equation supplemented 
by two further parameters acting as branch points of an elliptic curve. The emergence of this elliptic curve is 
rather mysterious at the level of Q3, which, unlike Q4, does not really warrant an elliptic parametrisation. 
Nonetheless, this curve naturally arises through what we consider to be the \textit{universal} parametrisation of 
all ABS equations (except perhaps Q4, which remains to be investigated), and which allows us to treat all equations 
in the list on the same footing (unlike the original parametrisation from \cite{ABS} where the parameters of the 
different equations in the list do not seem to be directly linked to each other). Thus, the $N$-soliton solution 
of Q3 can be written as a linear combination of four terms each of which contains as essential ingredient the 
$N$-soliton solution of the so-called NQC equation of \cite{NQC} with different values of the branch point parameters 
which enter in that equation. The $N$-soliton solutions of the other equations in the list, namely Q2, Q1, H3, H2, H1, 
follow by degeneration, and were derived in explicit form from the solutions of Q3. The present paper concentrated 
on an approach, using a Cauchy matrix representation of the basic objects, and which essentially was developed, as 
\textit{direct linearization approach} in the paper of the early 1980s. As a remarkable upshot a novel Miura transform 
between Q3 (almost at the top of the ABS list) and H1 (at the bottom of the list) played a crucial role in the 
mechanism behind the solutions. In the subsequent paper \cite{HieZhang} an alternative representation of the soliton 
solutions, in terms of Casorati determinants, is given, based on the bilinear forms of the ABS equations.  

We have not touched in this paper on continuum limits of the equations and the hierarchies of continuous equations 
associated with the lattice systems. This can obviously be done without problem. The most direct way of introducing 
continuum analogues is by simply replacing the discrete plane-wave factors $\rho_i$, which as functions of the 
discrete variables $n$,$m$ was given in \eqref{eq:rho} by exponentials, making the replacements:
$$ \rho_i=\left(\frac{p+k_i}{p-k_i}\right)^n\left(\frac{q+k_i}{q-k_i}\right)^m\ \rightarrow\  
e^{2k_ix+2k_i^3t}\qquad    {\rm and} \qquad  \digamma(a,b)\ \rightarrow\  e^{(a+b)x+(a^3+b^3)t}\   , $$ 
or by simply including the exponentials in the $\rho_i$ together with the discrete exponential factors (invoking 
once again multidimensional consistency). 
We leave the derivation of the corresponding PDEs as an exercise to the reader.    

\ack
The authors are grateful for the hospitality of the Isaac Newton Institute for Mathematical Sciences, Cambridge, where the present work was completed during the programme Discrete Integrable Systems (DIS).
JA was supported by the Australian Research Council (ARC) Centre of Excellence for Mathematics and Statistics of Complex Systems (MASCOS).

\end{document}